\documentclass[a4paper]{article}
\usepackage[latin1]{inputenc}
\usepackage{graphics}

\usepackage{amssymb}
\usepackage{amsmath}
\usepackage[thmmarks, amsmath]{ntheorem}
\usepackage[all]{xy}

\def\A{{\mathcal A}}
\def\B{{\mathcal B}}

\def\D{{\mathcal D}}
\def\G{{\mathcal G}}
\def\H{{\mathcal H}}
\def\J{{\mathcal J}}
\def\K{{\mathcal K}}
\def\M{{\mathcal M}}
\def\S{{\mathcal S}}

\def\RR{{\mathbb R}}
\def\CC{{\mathbb C}}

\def\ZZ{{\mathbb Z}}

\def\HH{{\mathbb H}}
\def\BB{{\mathbb B}}
\def\FF{{\mathbb F}}
\def\AA{{\mathbb A}}
\def\ext{{\rm ext}}
\def\Ad{{\rm Ad}}

\def\reel{{\rm Re}}

\def\End{\mbox{\rm End}}

\def\Aut{\mbox{\rm Aut}}

\def\bra{\langle}
\def\ket{\rangle}
\def\id{{\rm Id}}
\def\tr{\mbox{\rm Tr}}
\def\ker{\mbox{\rm Ker}}

\def\diag{{\rm diag}}

\def\bea{\begin{eqnarray}}
\def\eea{\end{eqnarray}}

\def\be{\begin{equation}}
\def\ee{\end{equation}}

\def\bat{\bar}

\newenvironment{rem}[1][{}]{\smallbreak \noindent  {\bf Remark #1}\small }

\newcommand*\colvec[1]{\begin{pmatrix}#1\end{pmatrix}}

\newtheorem{definition}{Definition}
\newtheorem{lemma}{Lemma}

\newtheorem{propo}{Proposition}
\theoremstyle{nonumberplain}
\theorembodyfont{\normalfont}
\theoremseparator{:}
\theoremsymbol{$\P$}

\newtheorem{demo}{Proof}
\begin{document}
\title{A $U(1)_{B-L}$-extension of the Standard Model from Noncommutative Geometry}
\author{Fabien Besnard}
\maketitle
\begin{abstract} We derive a $U(1)_{B-L}$-extension of the Standard Model from a generalized Connes-Lott model with algebra $\CC\oplus\CC\oplus \HH\oplus M_3(\CC)$. This generalization includes the Lorentzian signature, the presence of a real structure, and a weakening of the order $1$ condition. In addition to the SM fields, the model contains a $Z_{B-L}'$ boson and a complex scalar field $\sigma$ which spontaneously breaks the new symmetry.  This model is the smallest one which contains the SM fields and is compatible with both the Connes-Lott theory and the algebraic background framework. 
\end{abstract}

\section{Introduction}
Particle physics has for long been in search of a unifying principle. With the non-detection of proton decay or supersymmetric partners, it is not an overstatement  that GUT and String theory inspired models  are now facing a crisis. However, a few physicists and mathematicians have been developping another research program, known as Noncommutative Geometry, around the deep ideas set forth by Alain Connes since the 90's (see \cite{CVSsurvey} for a recent survey of the historical development of this program). Let us summarize its most salient features:

\begin{itemize}
\item It is based on the notion of real, even, \emph{spectral triples}: multiplets containing an algebra $\A$, a Hilbert space $\H$, a Dirac operator $D$, a chirality operator $\chi$ and a real structure $J$. They can be seen as the dual objects to virtual noncommutative Riemannian spin manifolds \cite{reconst}.
\item A particle physics model is obtained in 3 steps:
	\begin{enumerate}
	\item A finite-dimensional spectral triple $\S_F$ is chosen and tensorized with the canonical Spectral Triple of the manifold $M$, defining a virtual product $M\times F$, known as an almost-commutative manifold.
	\item The (bosonic) configuration space of the model is defined to be the space of \emph{fluctuated} Dirac operators, of the form $D_\omega=D+\omega+J\omega J^{-1}$ where $\omega$ is a noncommutative $1$-form.  
	\item A bosonic action functional $S_b(D_\omega)$ is defined. It is supplemented by a fermionic action of the form  $S_f(D_\omega,\Psi)=(\Psi,D_\omega \Psi)$, where $(.,.)$ is a suitable hermitian form.
	\end{enumerate}
	\item One thus obtain a classical field theory on an almost-commutative manifold.
\end{itemize}

The approach is conceptually satisfactory for several reasons:

\begin{description}
\item[a)] The Higgs  and  gauge sectors are unified: the space of noncommutative $1$-forms on an almost-commutative manifold naturally decomposes into two parts, one which can be identified with gauge fields and the other with Higgs fields.
\item[b)] Model building in NCG is far more constrained than in usual gauge theory, essentially because one starts with algebras instead of groups.
\item[c)]\label{point3} It is possible to find a finite triple $\S_F$ so that the configuration space contains all the fields of the Standard Model.
\item[d)] It is possible to find an action of Yang-Mills type, i.e. given by the norm of curvature of the noncommutative $1$-form $\omega$, which reproduces all the bosonic terms of the SM \cite{conneslott}. In particular the Higgs and gauge terms  have the same origin: all the bosonic fields of the SM are unified.  
\item[e)] It is possible to find an action, the Spectral Action, which depends on $\omega$ only through $D_\omega$, and which yields the Einstein-Hilbert action in addition to the SM terms.
\item[f)] There are less free parameters entering the action as in the usual SM. This yields to predictions at high energy (among which, the same prediction on gauge coupling as in GUT).
\end{description}

It is also very striking that all the intricacies of the SM, such as a the Higgs mechanism, neutrino oscillations and see-saw mechanism do not have to be added by hand: they ultimately follow from the form of the finite Dirac operator $D_F$ which is allowed by general principles having a geometric origin. This can be illustrated by the case of neutrino oscillations: in the first models they did not appear because some matrix elements of $D_F$ were set to zero by hand to match the physics known at the time. Now that neutrino oscillations are known, the finite Dirac is not tinkered anymore and the theory   looks all the better for it.  Moreover, the said matrix elements are constrained by the axiioms of spectral triples  to have a symmetry which is precisely the one which allows for the usual see-saw mechanism.  It must be said, however, that in the current models $D_F$ still does not have the most general possible form. Some elements are still set to zero by hand (though there exist certain algebraic prescriptions which seek to explain these zeroes, such as the massless photon condition \cite{connesmarcolli} or the second-order condition \cite{BF1}, \cite{BBB}). Yet, the noncommutative sky is not free of clouds. The most obvious problems are the following:

\begin{enumerate}
\item {\bf The first signature problem.} The theory of spectral triples is fundamentally Euclidean. 
\item {\bf The fermion doubling problem.} The definition of the finite Hilbert space includes degrees of freedom, which are needed to obtain the correct representation of the gauge group, but end up multiplying by four the dimension of the space of fermions fields.
\item {\bf The Higgs mass problem.}\label{higgsproblem} Under the big desert hypothesis, the Spectral Action can be used to predict the mass of the Higgs boson, but it turns out to be about $40 \%$ too high.
\item {\bf The second signature problem.} The Spectral Action is only defined in a Euclidean context.
\item {\bf The unimodularity problem.} In  {\bf c)} above, the configuration space unfortunately contains an additional field, a $U(1)_X$-boson, which is anomalous. One has to require an \emph{unimodularity condition} in order to remove it along with the extra $U(1)$-symmetry.
\end{enumerate}

A lot of work \cite{moretti,stro,PS,franco,vddgpr,vdd,part1,SST2,thesenadir,bbb2,doppler,algbackpart1,algbackpart2} has been devoted to the first signature problem over the years. Though no complete formalism is universally accepted yet, partial solutions, and a working replacement for spectral triples in non-Euclidean signature exist.

The fermion doubling problem has been pointed out in \cite{LMMS}. It is maybe the best understood by now. Two independent solutions have been given \cite{Barrett, SMmix}, which, quite strikingly, are only available in the precise KO-dimension which allows for neutrino mixing terms in the finite Dirac. Interesting links with the first signature problem have been found \cite{AKL}. It was also shown that Barrett's solution is unique under a natural symmetry principle, and is equivalent to a simple modification of the fermionic action \cite{barrettunique}. 

The Higgs mass problem is probably the less severe of all for the noncommutative geometry program, since it relies on the big desert hypothesis, which is not exactly compelling. However, it has been found \cite{resilience} that the addition of a   scalar field $\sigma$ to the model, already known to stabilize the electroweak vacuum \cite{EEGLS,lebedev} and for being a good dark matter candidate \cite{chentang}, can push the prediction down to a value very close to that obtained at the LHC. In \cite{resilience}, $\sigma$ is obtained by turning the neutrino mixing part of the finite Dirac into a field,  which is not natural since this part remains constant under fluctuations. Since then,  two  approaches to this question have emerged:  Boyle-Farnsworth theory (\cite{BF1}, \cite{BF2}), and  twisted Spectral Triples theory (\cite{DLM}, \cite{DLMproc}, \cite{DFLM}). Both require  important modifications to the usual formalism of NCG.

No solution for the two last problems have been proposed yet\footnote{The unimodularity problem does not show up when complex algebras are used \cite{LazzariniSchucker}, but the Standard Model is unfortunately formulated with real algebras.}. However, the second signature problem can be completely by-passed at the price of giving up on unifying gravity with the other forces. Indeed, the Connes-Lott action can be extended to the Lorentzian noncommutative SM with no problem \cite{thesenadir}. We will explain this idea in the present paper, in a form more suitable for generalizations.
 
In addition to the above problems, there is another, maybe less visible, one:

\begin{enumerate}\setcounter{enumi}{5}
\item\label{background} {\bf The background problem.} What are exactly the background structures in NCG ?  
\end{enumerate}

In particular, one can wonder if the Dirac operator around which we fluctuate is a background structure. If it is, then diffeomorphism invariance is broken. If it's not, then one can guess that fluctuated Dirac are just a part of a larger configuration space. Another way to consider the question is to ask about the automorphism group of a spectral triple. It is generally defined to be the set of unitary operators which commute with the real structure and the chirality, and stabilize the algebra. If one stops there and look at  the canonical spectral triple of a manifold, then one finds more examples than just diffeomorphisms and local Lorentz transformations. If we also require the commutation with the Dirac, we find only isometries. This is too many or too few. In \cite{algbackpart1}, we argued that the solution is leave the Dirac operator away from the background and include instead the bimodule of noncommutative $1$-forms. The automorphisms of this new structure, called \emph{algebraic background}, turn out to be exactly what one expects, diffeomorphism and local Lorentz transformations, in the manifold case. In the algebraic background framework, the bosonic configuration space is no longer restricted to the fluctuations around a given Dirac, it is the space of all the Dirac operators compatible with the bimodule of $1$-forms, complying fully to the spirit of Kaluza-Klein theories. However, there is a twist when we apply this idea to the manifold case: we obtain more than just metrics. The additional fields fields are called \emph{centralizing} because they commute with the algebra. The configuration space is thus the direct sum $\Delta\oplus {\mathcal Z}$ of Dirac operators associated with metrics and centralizing fields. Remarkably, these two parts are separately invariant under the symmetry group, and it is thus possible to project to $\Delta$. The application of these ideas to the SM leads to an interesting conclusion: the SM alone cannot be obtained in this framework ! More precisely, using the usual SM algebra and space of noncommutative $1$-forms, it is found that the automorphism group of the algebraic background has an extra $U(1)$ factor which can be identified with a (gauged) $B-L$-symmetry group. The configuration space contains, in addition to the SM fields, a vector boson associated to the $B-L$ symmetry, another one associated to the anomalous $U(1)$ part of the gauge group (the one which is removed by unimodularity), the $\sigma$ scalar boson, and flavour changing fields. The latter can be eliminated by a gauge-invariant projection. This is already an interesting result since it shows that the resolution of problem \ref{background} provided by the algebraic background framework points towards the $U(1)_{B-L}$-extension of the SM which has long been attracting physicists' attention, and makes the $\sigma$-field appear naturally. Alas this model is not entirely satisfactory. Indeed, while it is perfectly possible to use the Spectral Action in the Euclidean signature, the Connes-Lott action cannot be used in the Lorentzian case since it only applies to $1$-forms, and neither the $\sigma$-field nor the $Z'_{B-L}$ boson are $1$-forms.

For this reason, we will consider in this paper a simple modification of the model which consists only of extending the algebra by a factor of $\CC$. Doing so, the unification of the Higgs and gauge sector is recovered and all fields (except gravity) become $1$-forms subject to a common action principle. Abelian extensions of the Noncommutative Standard Model have been studied in the past, in particular in \cite{KrajPris}. However, they were always constrained by the order $1$ condition, while our model, which is a sub-model of Pati-Salam, does not satisfy this condition. Nevertheless, it satisfies a weaker condition which proves to be sufficient to apply the generalized Connes-Lott formalism. Hence our model is new as far as NCG is concerned\footnote{It must be stressed, though, that the exact same field content has been considered in  \cite{theseShane}, in the context of Non-associative Geometry, and with a Euclidean signature. For a summary of this approach, see \cite{BF1}.}. On the other hand $U(1)$ extensions of the SM with an additional  Higgs to break the new symmetry is one of the best motivated BSM model and has been  extensively studied from a phenomenological point of view \cite{holdom, ADH, CRM, BMP, BMP2, BBMST, IOO, boylecosmo}. See also the very useful pedagogical introduction \cite{salvioni}.

Hence we see that algebraic backgrounds, initially formulated to solve  problem \ref{background} also provide a solution to problem \ref{higgsproblem}. 


The present paper is organized as follows. In section 2 we recall the main points of the algebraic background framework, and give the example of the manifold and SM backgrounds. We also define the $J$-symmetrization of a background. In section 3 we recall Connes' theory of noncommutative 1-forms, and explain how to compute their curvature in the presence of a real structure by embedding them into the $J$-symmetrized background.  Section 4 is devoted to the different bosonic configuration spaces which will be used in the paper. In section 5 we show that a Connes-Lott theory of the SM in the Lorentzian signature is perfectly well-defined, though it is inconsistent with the algebraic background framework. To solve this inconsistency, we extend the finite background in section 6, and compute the bosonic action in section 7. Section 8 helps identifying the fields written in the NCG  way with those known by physicists, and derive some relations between particle masses and couplings. Section 9 is devoted to the fermionic action, and section 10 offers a conclusion. The cumbersome calculations of junk and $J$-symmetrized junk bimodules are given in the appendices.

In the whole text we use the following general notations: a diagonal or block-diagonal matrix with diagonal entries $a,b,c,\ldots$ will be written $[a,b,c,\ldots]$, the complex conjugate of an object $A$ will be written $A^*$, the Hilbert adjoint $A^\dagger$, and the Krein adjoint $A^\times$.


\section{Algebraic backgrounds}
\subsection{General definitions}
In this paper we will use the general settings of algebraic backgrounds introduced in \cite{algbackpart1}, which we will now briefly review\footnote{In this paper, and contrarily to \cite{algbackpart1}, we will use the graded real structure $J$ instead of the real structure $C$ since this convention is more common in particle physics applications of NCG.}.

First, a \emph{pre-Krein space} is a vector space $\K$ equipped with a non-degenerate indefinite metric, which is decomposable into the direct sum $\K=\K_-\oplus \K_+$ of a positive and negative definite subspaces.  Such a decomposition is equivalent to   a fundamental symmetry $\eta$, which in turns defines a corresponding norm $\|.\|_\eta$. Note that in contrast with the case of Krein spaces, the $\eta$-norms need not be all equivalent \cite{doppler}.  

We will need to equip our pre-Krein spaces with more structures: a chirality and a real structure. This yields the following definition.

\begin{definition} A $\ZZ_2$-graded real pre-Krein space is a pre-Krein space $\K$ equipped with a linear operator $\chi$ (chirality) and an antilinear operator $J$ (graded real structure) such that
\be 
\chi^2=1,\quad J^2=  \epsilon,\quad J\chi=  \epsilon''\chi J,\quad J^\times  =  \kappa J,\quad \chi^\times =  \epsilon''  \kappa''\chi\label{kosigns}
\ee
where $\epsilon,\kappa,\epsilon'',\kappa''$ are signs (``KO-metric signs''). A fundamental symmetry $\eta$ is said to be compatible with $\chi$ and $J$ iff 
\be
\chi\eta=\epsilon''\kappa''\eta\chi\mbox{ and }J\eta=\epsilon\kappa \eta J\label{compsigns}
\ee
\end{definition}


We recall that  $\epsilon,\epsilon''$ are given in terms of $n$,  an integer modulo $8$ called the KO-dimension, by the formulas $\epsilon=(-1)^{n(n+2)\over 8}$, $\epsilon''=(-1)^{n/2}$, while $\kappa=(-1)^{m(m+2)\over 8}$, $\kappa''=(-1)^{m/2}$, where $m$ is another integer modulo $8$ called the metric dimension (for more details see \cite{bbb2}). For convenience the values of the signs $\epsilon,\epsilon'',\kappa,\kappa''$ in terms of $m,n$ are gathered in table \ref{tab1}.
\begin{table}[hbtp]
\begin{center}
\begin{tabular}{|c||c|c|c|c|}
\hline
m,n&0&2&4&6\\
\hline
$\kappa,\epsilon$&1&-1&-1&1\\
\hline
$\kappa'',\epsilon''$&1&-1&1&-1\\
\hline
\end{tabular}
\end{center}
\caption{ Signs $\epsilon,\epsilon'',\kappa,\kappa''$ in terms of $m,n$.}\label{tab1}
\end{table}

\begin{definition} Let   $A$ be a linear or anti-linear  operator on $\K$. Let $\|A\|_\eta$ be the operator norm of $A$ subordinated to the $\eta$-norm. We say that $A$ is \emph{universally bounded} if $\sup_\eta\|A\|_\eta<\infty$ where the supremum extends over all compatible fundamental symmetries.
\end{definition}
 
In particular $J$ and $\chi$ are universally bounded. Let $\B_u(\K)$ be the space of linear universally bounded operators on $\K$. It is clear that  $\B_u(\K)$ is a unital algebra and that \emph{the universal operator norm} 
\be 
\|A\|_u:=\sup_\eta\|A\|_\eta
\ee
defines a sub-multiplicative norm on it. Moreover, $A^\times$ is universally bounded if $A$ is, with the same universal norm, and for every compatible fundamental symmetry $\eta$, $(\B_u(\K),*_\eta,\|\ \|_\eta)$  is a pre-$C^*$-algebra.

The $\ZZ_2$-graded real pre-Krein space $\K$ can be decomposed into even and odd subspaces, $\K=\K_0\oplus \K_1$, which are the eigenspaces of $\chi$. An operator $A$ which commutes with $\chi$ will respect this decomposition and will be called \emph{even}. If $A$  anticommutes with $\chi$ it will exchange $\K_0$ and $\K_1$ and be called \emph{odd}. We also say that $A$ is  \emph{$J$-real} if it commutes with $J$, and \emph{$J$-imaginary} if it anticommutes with it. Note that if $\epsilon''\kappa''=1$ then $\chi^\times=\chi$ and this implies that $\K_0$ and $\K_1$ are orthogonal with respect to $(.,.)$. In this case we will say that the Krein product is even. On the contrary if $\epsilon''\kappa''=-1$, $\K_0$ and $\K_1$ are self-orthogonal ($\K_i=\K_i^\perp$) and we say that the Krein product is odd.

\begin{definition}\label{algbgd} An \emph{algebraic background} is a tuple ${\cal B}=(\A, \K, (.,.),\pi,\chi,J,\Omega^1)$  where:
\begin{enumerate}
\item$(\K,(.,.),\chi,J)$ is a $\ZZ_2$-graded real pre-Krein space, 
\item $\A$ is an algebra and $\pi$ is  a    representation of it on $\K$ by universally bounded operators,
\item the   chirality operator  $\chi$ commutes with $\pi(a)$ for all $a\in \A$,
\item the ``bimodule of 1-forms'' $\Omega^1$ is an  ${\cal A}$-bimodule of universally bounded operators on $\K$ such that for any $\omega\in \Omega^1$, $\omega\chi=-\chi\omega$.
\end{enumerate}
\end{definition}

The most important example is the canonical background over a spin semi-Riemannian manifold \cite{algbackpart1}. Let us describe here the only case we will need. Let $(M,g)$ be an open anti-Lorentzian manifold of dimension $4$ (signature $(+,-,-,-)$), which is space and time orientable and has a spin structure. In this case, as shown in \cite{geroch1}, there exists a global orthonormal tetrad $(e_a)_{0\le a \le 3}$. The spinor bundle will be ${\cal S}=M\times S$, where $S=\CC^4$ and the action of $TM$ on ${\cal S}$ is defined by the assignment $\gamma(e_a):=\gamma_a$, where gamma matrices   are chosen to be
\bea
\gamma^0=\gamma_0=\colvec{0&1_2\cr 1_2&0};\gamma^k=-\gamma_k=\colvec{0&-\sigma^k\cr \sigma^k&0}, k=1,2,3\cr
\mbox{with }\sigma^1=\colvec{0&1\cr 1&0}, \sigma^2=\colvec{0&-i\cr i&0}, \sigma^3=\colvec{1&0\cr 0&-1}
\eea
Note that $\gamma$ uniquely extends to the Clifford bundle. The chirality $\chi_M$ and charge conjugation $J_M$ are defined by $(\chi_M \Psi)(x)=\chi_S\Psi(x)$, $(J_M\Psi)(x)=J_S\Psi(x)$, where the ``local'' chirality and real structure $\chi_S$ and $J_S$ are defined as follows: $\chi_S$ is the multiplication by the matrix $\gamma_5=i\gamma^0\ldots\gamma^3=[I_2,-I_2]$, and  the real structure $J_S$ is $\psi\mapsto\gamma^2  \psi^*$, where $\psi^*$ is the complex conjugate of $\psi$ in the chosen basis. One can easily check that  $\chi_S^\times=-\chi_S$, $J_S$ anticommutes with gamma matrices and satisfies $J_S^2=1$, $J_S^\times=-J_S$. Let $\A_M$ be the algebra $\tilde {\cal C}^\infty_c(M,\RR):={\cal C}^\infty_c(M,\RR)\oplus \RR.1$, where $1$ is the constant function.    The pre-Krein space ${\cal K}_M$ is the space of smooth   spinor fields with compact support equipped with  the indefinite product:
\be
(\Psi,\Phi)=\int_M(\Psi(x),\Phi(x))_S{\rm vol}_g
\ee
where $(\phi,\psi)_S:=\phi^\dagger\gamma_0 \psi$ for any $\phi,\psi\in S$. The algebra $\A_M$ is represented on $\K_M$ by multiplication, i.e. $(\pi_M(f)\Psi)_x=f(x)\Psi_x$, with obvious notations. Finally, the bimodule of 1-forms $\Omega^1_M$ is generated by commutators $[D_M,\pi_M(f)]$ where  $D_M=i\gamma^\mu\nabla_\mu$, is the canonical Dirac operator. It is thus the space of fields of the form $if_\mu\gamma^\mu$, where $f_\mu$ is a smooth real function with compact support. The canonical algebraic background defined by all these data on the manifold is then
$$\B_M=(\A_M,\K_M,(.,.),\pi_M,\chi_M,J_M,\Omega^1_M).$$

An isomorphism between the algebraic backgrounds $\B=({\cal A},{\cal K},\ldots)$ and  ${\cal B}'=({\cal A}',{\cal K}',\ldots)$ is a Krein-unitary transformation $U$ such that $U\pi(\A)U^{-1}=\pi'(\A')$,  $UJU^{-1}=J'$, $U\chi U^{-1}=\chi'$ and $U\Omega^1U^{-1}=(\Omega^1)'$. An important particular case is the group $\Aut(\B_M)$ of automorphisms of the canonical background described above. It is  generated by two kinds of elements \cite{algbackpart1}:
\begin{enumerate}
\item Diffeomorphisms of the base, acting on spinor fields by pullback and rescaling according to: 
\be 
(U_\theta\Phi)_x=\sqrt{\frac{\rm vol_{\theta_*g}}{\rm vol_g}}\Phi_{\theta^{-1}x}
\ee
where $\theta$ is a diffeomorphism, $\Phi\in \K_M$,  and $x\in M$.
\item Spinomorphisms, i.e. local change of spin structure, acting by:
\be 
(U_\Sigma\Phi)_x=\Sigma_x\Phi_x 
\ee
where $\Sigma$ is a smooth map of $M$ to the neutral component of ${\rm Spin}(1,3)$. These can also be viewed as local Lorentz transformations lifted to the spin group.
\end{enumerate}
 
Let us return to the general case. We define the linear anti-automorphism  $A\mapsto A^o$ of $\End(\K)$ by:
\be 
A^o=JA^\times J^{-1}
\ee
There is thus a right representation of $\A$ on $\K$, defined by
\be 
\pi^o(a):=\pi(a)^o=J\pi(a)^\times J^{-1}\label{opprep}
\ee
A background will be said to satisfy the \emph{order $0$ condition} ($C_0$) if for all $a,b\in\A$ one has
\be 
[\pi(a)^o,\pi(b)]=0\label{C0}
\ee 
It will be said to satisfy the \emph{order $1$ condition} ($C_1$) if for all $a\in \A$ and $\omega\in \Omega^1$, one has
\be 
[\pi(a)^o,\omega]=0\label{C1}
\ee 
This condition turns out to be too restrictive for our purpose and we will have to replace it with the \emph{weak order $1$ condition} (weak $C_1$): for all invertible $a\in \A$, 
\be 
\pi(a)^o\Omega^1\pi(a^{-1})^o=\Omega^1\label{weakC1}
\ee
It is worthy of  note that while the canonical background over a manifold satisfies $C_1$, the canonical background over a finite graph only satisfies weak $C_1$ (see \cite{SST2} for the definition). All the backgrounds considered in this paper will satisfy $C_0$ and at least  weak $C_1$. Moreover, if $\A$ is a $*$-algebra (resp. pre-$C^*$-algebra), and $\pi$ is a $*$-representation ($\pi(a^*)=\pi(a)^\times$), then $\B$ will be called a $*$- (resp. pre-$C^*$-) algebraic backgrounds. All the backgrounds in this paper will be pre-$C^*$. In that case, one can define the unitary group of $\A$: $U(\A)=\{a\in \A|uu^*=1\}$. The image of the unitary elements of $\A$ under $\pi$ are not automorphisms of the background in general, since they do not commute with $J$. Instead, for $a\in U(\A)$ one defines 
\be 
\Upsilon(a)=\pi(a)J\pi(a)J^{-1}=\pi(a)\pi^o(a^{-1})
\ee
Thanks to $C_0$ and weak $C_1$, one has $\Upsilon(a)\in \Aut(\B)$, so that $\G_A:=\Upsilon(U(\A))$ is a subgroup of $\Aut(\B)$ called \emph{the gauge group}, using the notation and terminology of \cite{VS}.

We will also make use of the following notion: let $\B$ be a background satisfying $C_0$. Then the \emph{$J$-symmetrized background} $\hat\B$ is obtained by replacing:
\begin{itemize}
\item $\A$ with the algebra $\hat \A$ generated by $\pi(\A)$ and $\pi(\A)^o$,
\item $\pi$ with $\hat\pi=\id$,
\item $\Omega^1$ with $\hat \Omega^1$, which the $\hat \A$-bimodule generated by $\Omega^1$ and $(\Omega^1)^o$,
\end{itemize} 
all the other pieces of data remaining unchanged. Note that, using $C_0$, $\hat \A$ is the image of the  envelopping algebra $\A\otimes \A^o$ under $a\otimes b^o\mapsto\pi(a)\pi(b)^o$.

\begin{definition}\label{defdirac}   A \emph{Dirac operator} on the algebraic background $\B$ is  a symmetric operator on $\K$ which is odd, $J$-real, and such that for all $a\in \A$, $[D,\pi(a)]\in \Omega^1$. It is said to be \emph{regular} if $\Omega^1$ is generated as an $\A$-bimodule by the commutators $[D,\pi(a)]$, $a\in\A$. 
\end{definition}
The space of all   Dirac operators is called the \emph{configuration space} of $\B$, and written ${\cal D}_\B$. Note that $\D_\B\subset \D_{\hat\B}$ and that this inclusion preserves regular elements.

\subsection{The algebraic background of the Lorentzian Noncommutative Standard Model}
The Lorentzian NCSM is based on an almost-commutative background $\B=\B_M\hat\otimes \B_F$, where $\B_M$ is the canonical background over a Lorentzian $4$-manifold, and  $\B_F:=(\A_F,\ldots,\Omega^1_F)$ is a finite background which we are now going to describe. In order to define ${\cal K}_F$, we first define a space ${\cal K}_0=(\CC^2\oplus \CC^2\otimes \CC^3_c)\otimes \CC^N_g$. The integer $N$ is the number of generations, which is here arbitrary. The canonical basis of ${\cal K}_0$ is labeled as follows: the basis of the first $\CC^2$ is $(\nu,e)$, the basis of the second one is $(u,d)$, the basis of the color $\CC^3$ is $(r,b,g)$, and the basis of the generation $\CC^3$ is $(e_1,e_2,e_3)$. We also introduce $I\simeq \CC^4$,   the vector space generated by the four symbols $R,L,{\bar R},{\bar L}$. We decompose ${\cal K}_F$ as the direct sum ${\cal K}_F={\cal K}_R\oplus {\cal K}_L\oplus {\cal K}_{\bar R}\oplus {\cal K}_{\bar L}$, where ${\cal K}_\sigma={\cal K}_0\otimes \sigma$. A vector of the form $\psi\otimes \sigma\in{\cal K}_\sigma$ will often be written $\psi_\sigma$. We will often  see operators on ${\cal K}_F$ as block $4\times 4$-matrices with entries in $\End({\cal K}_0)$. Another useful piece of notation is the following. Identify ${\cal K}_0$ with $\CC^2\otimes \CC^4\otimes \CC^N_g$ (seeing lepton as a fourth color, as in the Pati-Salam model). Then for any element $a$ of $M_2(\CC)$ we write $\tilde a=a\otimes 1\otimes 1$. Returning to the first decomposition of ${\cal K}_0$ we have $\tilde a=(a\oplus a\otimes 1)\otimes 1$. It will also be convenient to introduce a notation for Krein selfadjoint projectors on subspaces of $\K_F$ corresponding to different types of particles. The general notation will be $p_s$ where $s$ is a particle symbol. For instance,  $p_{q}$, the projector on quark space is $p_{q}=[0\oplus 1_2\otimes 1_3,0\oplus  1_2\otimes 1_3,0,0]\otimes 1_N$. The meaning will always be clear by the context.

We denote by $\bra .,.\ket$ the canonical scalar product on ${\cal K}_F$.  We define the Krein product with  the following fundamental symmetry $\eta_F$, which we also call the internal metric:
\begin{equation}
\eta_F=[1,-1,s,-s]
\end{equation}
where $s=\pm 1$ and where $1$ means the identity of ${\cal K}_0$. We are thus considering two cases. In order to recover the correct fermionic action we must take $s=1$ if we consider commuting fermion variables, and $s=-1$ if we consider them to be anti-commuting, as in traditional QFT \cite{thesenadir}.

The finite algebra is $\A_F=\CC\oplus \HH\oplus M_3(\CC)$. Its representation  is:
\be
\pi_F(a)(\lambda,q,m)=[\tilde q_\lambda,\tilde q,\lambda\otimes 1_N\oplus 1_2\otimes m\otimes 1_N,\lambda\otimes 1_3\oplus m\otimes 1_N]
\ee
where $q_\lambda=\colvec{\lambda&0\cr 0&\lambda^*}$ for any $\lambda\in \CC$. The finite chirality and real structure are
\bea
\chi_F&=&[1,-1,-1,1]\\
J_F(\phi\otimes \sigma)&=& \phi^*\otimes\bar\sigma
\eea
for all $\phi\in{\cal K}_0$ and $\sigma=R,L,\bar R,\bar L$, with the convention that  $\bar{\bar \sigma}=-\sigma$. One can check that $\epsilon_F=-1$, $\epsilon_F''=-1$, $\kappa_F''=-1$, $\kappa_F=-s$, so that the KO-metric dimension pair $(m_F,n_F)$ is $(2,2)$ if $s=1$ and $(6,2)$ if $s=-1$.

The bimodule $\Omega^1_F$ is defined as follows. Let  $D_F$  be:
\be
D_F=\colvec{0&-Y_0^\dagger&sM_0^\dagger&0\cr 
Y_0&0&0&0\cr
M_0&0&0&-Y_0^T\cr 
0&0&Y^*_0&0}
\ee
where 
\be
Y_0=\colvec{Y_\nu&0\cr 0& Y_e}\oplus \colvec{ 1_3\otimes Y_u&0\cr 0&1_3\otimes Y_d}
\ee
and 
\be 
M_0=\colvec{1&0\cr 0&0}\otimes m_0:=p_\nu\otimes m_0
\ee
where $m_0\in M_N(\CC)$ is non-vanishing and satisfies $m_0^T=-sm_0$. We ask $D_F$ to be a regular Dirac operator. It follows easily that 
\be
\Omega^1_F=\{\colvec{0&Y_0^\dagger \tilde q_1&0&0\cr \tilde q_2 Y_0&0&0&0\cr 0&0&0&0\cr 0&0&0&0
}|q_1,q_2\in\HH\} \label{finite1forms}
\ee
In the sequel, the \emph{mass matrices} $Y_\nu Y_\nu^\dagger:=Y_\nu Y_\nu^\dagger,\ldots, Y_d Y_d^\dagger:=Y_dY_d^\dagger$ and $m_0m_0^\dagger:=m_0m_0^\dagger$ will play an important role. In particular, we will need to assume the  following \emph{genericity hypothesis}: 
\begin{enumerate}
\item $Y_0$ is invertible,
\item any element of $M_2(\CC)$ (resp. $M_3(\CC)$) commuting with $Y_\nu Y_\nu^\dagger$ and $Y_e Y_e^\dagger$ (resp. $Y_u Y_u^\dagger$ and $Y_d Y_d^\dagger$) is scalar.
\end{enumerate}

\begin{rem} The choice of $D_F$ is severely constrained by the axioms of Noncommutative Geometry (see \cite{connesmarcolli}, \cite{BF1} for the Euclidean case, and \cite{thesenadir} for the indefinite case). Note also that for the usual   see-saw mechanism to occur, a symmetric $m_0$ is needed, which implies $s=-1$.
\end{rem}
Let us now look at the gauge group  ${\cal G}_{A_F}$. Its elements have the form $U=[A,B,A^*,B^*]\otimes 1_N$ where 
\bea 
A&=&(\colvec{1&0\cr 0&e^{-2i\theta}}\oplus \colvec{e^{i\theta}&0\cr 0&e^{-i\theta}}\otimes \bar m)\cr
B&=&(e^{-i\theta}q\oplus q\otimes \bar m)\otimes 1_N\label{gaugegroup}
\eea
with $q,m$  unitary quaternions and $3\times 3$ matrices, and $\theta$ a  real number. The \emph{unimodular gauge group} $S\G_{\A_F}$ is the subgroup of $\G_{\A_F}$ defined by the condition $\det(\pi_F(u))=1$. Its elements are of the form 
\bea
A&=&\colvec{1&0\cr 0&e^{-2i\theta}}\oplus \colvec{e^{4i\theta/3}&0\cr 0&e^{-2i\theta/3}}\otimes \bar g \cr
B&=&qe^{-i\theta}\oplus qe^{i\theta/3}\otimes \bar g\label{genunimod}
\eea
with $g\in SU(3)$. For more details see \cite{thesenadir}. Since $\B_F$ satisfies $C_1$, $\G_{\A_F}$ is a subgroup of $\Aut(\B_F)$. The latter also contains $U(1)_{B-L}$, the group of automorphisms of the form:
\be 
g_{B-L}(t)=[A(t),A(t),{A(t)}^*,{A(t)}^*]\otimes 1_N,\mbox{ with }A(t)=e^{-it}1_2\oplus e^{it\over 3}1_2\otimes 1_3.
\ee
Using the genericity hypothesis, one can prove that $\Aut(\B_F)$ is generated by $\G_{\A_F}$ and $U(1)_{B-L}$ \cite{algbackpart2}.

The total algebraic background $\B=(\A,\ldots,\Omega^1)$ of the Standard Model is the graded tensor product of $\B_M$ with $\B_F$. According to the general rules exposed in \cite{algbackpart2}, we thus have $\A=\tilde{\cal C}^\infty_c(M,\A_F)$, $\K=\Gamma^\infty_c(S\otimes \K_F)$, $\chi=\chi_M\otimes \chi_F$, $\pi=\pi_M\otimes \pi_F$, $J=J_M\chi_M\hat\otimes J_F\chi_F=J_M\otimes J_F\chi_F$, and the bimodule of 1-forms is obtained from \eqref{decomp1form}. The less obvious part is  the Krein product on ${\cal K}={\cal K}_M\otimes {\cal K}_F$, which is defined by the integral over $M$ of the following ``local Krein product'' defined on $S\otimes {\cal K}_F$:
\begin{equation}
(\psi\otimes\phi,\psi'\otimes\phi')=(\psi,\psi')_{S}\bra\phi,\omega\phi' \ket
\end{equation}
where the ``effective internal metric'' $\omega$ is 
\begin{equation}
\omega=\chi_F\eta_F=[1,1,-s,-s]
\end{equation}
For the explanation of this strange-looking rule, see  \cite{bbb2}. Let us turn to the group $\Aut(\B)$.  It is proven\footnote{The proof uses the hypothesis that $H_1(\pi_1(M),\ZZ)=\{1\}$, which is obviously satisfied here. Counter-examples exist without this hypothesis.} in \cite{algbackpart2} that it  is generated by $\Aut(\B_M)\hat\otimes \id_{\K_F}$ as well as 
\begin{itemize}
\item \emph{local gauge symmetries}, i.e. elements of the gauge group $\G_\A$, which are of the form $\Upsilon(u)$, where $u\in U(\A)$  is a field with values in $U(\A_F)$,
\item \emph{local $B-L$ symmetries}, of the form  $U_\varphi$, where $\varphi$ is a real field, and $U_\varphi(\Phi\otimes \psi)_x=\Phi_x\otimes g_{B-L}(\varphi(x))\psi$, for all  $\Phi\in \K_M$ and $\psi\in \K_F$. 
\end{itemize}

We end this section with some useful formulas. Let $A\in {\rm End}(\K_M)$ and $B\in{\rm End}(\K_F)$. Then:
\bea
J(A\hat\otimes B)J^{-1}&=&J_MAJ_M^{-1}\hat\otimes J_F BJ_F^{-1}\cr
(A\hat\otimes B)^\times&=&(-1)^{|A||B|}A^\times\hat\otimes B^\times\cr
(A\hat\otimes B)^o&=&(-1)^{|A||B|}A^o\hat\otimes B^o\label{formulesympa} 
\eea
and
$$J_F\colvec{A&B\cr C&D}J_F^{-1}=\colvec{D^*&-C^*\cr -B^*&A^*}$$

\section{The curvature of noncommutative forms}\label{curvform}
In Connes-Lott gauge theory we will have to compute the curvature of noncommutative 1-forms, and in this section we briefly recall how this is done. The general theory of  noncommutative forms of all degrees in Euclidean NCG can be found in \cite{redbook}, chap. VI. Here we focus only on forms on degree $0,1,2$ and on the small modifications introduced by the presence of a real structure and the replacement of a Hilbert space by a pre-Krein space. We defines forms directly in representation.

Let $\B=(\A,\ldots,\Omega^1)$ be an algebraic background. Any    Dirac operator $D$ for $\B$ defines a derivation $d_D : \pi(\A)\rightarrow \Omega^1$ by $d_D(\pi(a))=[D,\pi(a)]$. We seek to extend this derivation to the $\A$-bimodule $\Omega^1$.  For this, let $\J^1_D$ be the subspace of elements of $\End(\K)$ of the form
\be 
\sum_j [D,\pi(a_j)][D,\pi(b_j)],\mbox{ with } a_j,b_j\in\A\mbox{ s.t. }\sum_j \pi(a_j)[D,\pi(b_j)]=0
\ee
These elements are called \emph{junk $2$-forms}, and $\J^1_D$ is an $\A$-bimodule.

If $D$ is regular, any $1$-form $\omega$ can be written as $\omega=\sum_j\pi(a_j)[D,\pi(b_j)]$, though not in a unique way. However, the expression
\be 
d_D(\omega)=\sum_j[D,\pi(a_j)][D,\pi(b_j)]+\J^1_D
\ee
is a well-defined class modulo junk forms and satisfies the graded Leibniz rule, i.e. $d_D(\pi(a)\omega)=d_D(\pi(a))\omega+\pi(a)d_D(\omega)$ and  $d_D( \omega\pi(a))=d_D(\omega)\pi(a)-\omega d_D(\pi(a))$, as well as $d_D^2(\pi(a))=0$, for all $a\in\A$, $\omega\in \Omega^1$. If $\B$ is a $*$-background, we have the following properties with respect to the involution: $d_D(\pi(a^*))=-d_D(\pi(a))^\times$, and $d_D(\omega^\times)=(d_D\omega)^\times$. All these identities hold modulo junk forms. Finally, the \emph{curvature} of the $1$-form $\omega$ is defined to be
\be 
\rho_D(\omega):=d_D\omega+\omega^2
\ee
Each invertible element $u\in \A$ defines a \emph{gauge transformation} on $\Omega^1$ which is the map
\be 
\omega\mapsto\omega^u:=u\omega u^{-1}+u[D,u^{-1}]\label{gaugetrans}
\ee
Then for all $1$-form $\omega$, the curvature $\rho_D(\omega)$ is gauge-covariant, i.e.
\be 
\rho_D(\omega^u)=u\rho_D(\omega)u^{-1}\label{gaugecov}
\ee
Note that we do not need the order $0$ or $1$ condition to derive this result, so that we can in particular apply it on the $J$-symmetrized background $\hat \B$.

Let us now consider the case of an almost-commutative $*$-algebraic background $\B=\B_M\hat\otimes \B_F$. The bimodule of $1$-forms of $\B$ is obtained from that of $\B_M$ and $\B_F$ by
\be
\Omega^1=\Omega^1_{M}\otimes \pi_F(\A_F)\oplus \pi_M(\A_M)\otimes \Omega^1_{F}.\label{decomp1form}
\ee
This decomposition is such that if $D_M$ and $D_F$  are regular Dirac operators on $\B_M$ and $\B_F$ respectively, then $D=D_M\hat\otimes 1+1\hat\otimes D_F$ is a regular Dirac operator on $\B$. Moreover, for such a $D$, there is a similar decomposition of the junk $2$-forms  \cite{kalau}, \cite{martinlow}, \cite{BBB}, \cite{thesenadir}:
\bea
\J^1_{D}&=&\J^1_{D_M}\otimes \pi_F(\A_F)\oplus \pi_M(\A_M)\otimes \J^1_{D_F}\cr
&=&\tilde{\cal C}^\infty_c(M,\RR)\otimes \pi_F(\A_F)\oplus \tilde{\cal C}^\infty_c(M,\RR)\otimes \J^1_{D_F}\cr
&=&\tilde{\cal C}^\infty_c(M,\pi_F(\A_F)+\J^1_{D_F})
\eea
where in the second line we used the fact that a regular Dirac on $\B_M$ is the canonical Dirac plus a zero-order term \cite{algbackpart1}.

Now let $\omega\in\Omega^1$. Let us compute the curvature of $\omega+\omega^o\in \hat\Omega^1$. We express the result as a lemma for future reference.

\begin{lemma}\label{curvomegaz} We have $\rho_D(\omega+\omega^o)=\rho_D(\omega)+\rho_D(\omega)^o+\{\omega,\omega^o\}$. Moreover, if $\B$ satisfies $C_1$, then $\{\omega,\omega^o\}=0$.
\end{lemma}
\begin{demo}
First we note that if $\omega=a[D,b]$ then $\omega^o=[D,b]^oa^o=-[D,b^o]a^o=b^o[D,a^o]-[D,b^oa^o]$, from which it follows that $d_D(\omega^o)=[D,b^o][D,a^o]=(d_D\omega)^o$, modulo $\hat\J^1_D$. By linearity we thus obtain that 
\bea
\rho_D(\omega+\omega^o)&=&d_D\omega+(d_D\omega)^o+\omega^2+(\omega^2)^o+\omega\omega^o+\omega^o\omega\cr
&=&\rho_D(\omega)+\rho_D(\omega)^o+\{\omega,\omega^o\}
\eea
Now if $\B$ satisfies $C_1$, we obtain from $a^o[D,b]=[D,b]a^0$ that $a^0[D,b]+b[D,a^o]-[D,ba^o]=0$. Hence $[D,a^o][D,b]+[D,b][D,a^o]$ is in the junk. Thus $1$-forms and opposite $1$-forms anticommute up to junk. 
\end{demo}
%
%
 
%
For any two  operators $A_i$, $i=1,2$ on a finite-dimensional Krein space $W$, we can define their \emph{Krein-Schmidt product} by
\be
(A_1,A_2)=\tr(A_1^\times A_2)
\ee
The adjoint of an operator $T$ on $\End(W)$ with respect to this product will be denoted by $T^\times$. This is consistent since $L_A^\times=L_{A^\times}$ and $R_A^\times=R_{A^\times}$ where $L_A$ and $R_A$ are the left and right multiplication by $A\in \End(W)$.  The Krein-Schmidt product extends by integration to the endomorphisms of the pre-Krein space of an almost-commutative background which   are of the special form $(A_i\Psi)_x=A_i(x)\Psi_x$ with $A_i(x)\in \End(S\otimes \K_F)$, where $S$ is the space of Dirac spinors. Hence we  define the  \emph{integral Krein-Schmidt product} by
\be
(A_1,A_2):=\int_M \tr(A_1^\times(x) A_2(x)){\rm vol_g}
\ee
In the sequel we will not remind whether the Krein-Schmidt products are  integral or not, since this will be clear by the context.  We can also define the \emph{real Krein-Schmidt products} by $(.,.)_\RR:=\reel (.,.)$. It is a symmetric bilinear form on $\End(S\otimes \K_F)$ considered as a real space.    Since $\Omega^1$, $\hat\Omega^1$, etc. are all real vector subspaces of $\End(\K)$, this will be the privileged tool in what follows.  Note that the real Krein-Schmidt product on $\K_F$ satisfies the property
\be 
(A_1^o,A_2)_\RR=(A_1,A_2^o)_\RR\label{eq94}
\ee

Now, and for the rest of this section, we suppose that the real Krein-Schmidt product is non-degenerate on the subspace $V=\pi_F(\A_F)\oplus\J^1_{D_F}$ of $\End(S\otimes \K_F)$. Then $V\cap V^\perp=\{0\}$ and there is a well-defined projection operator $P$ (which depends on $D_F$) on $V^\perp$ with respect to $V$, which satisfies\footnote{Here $\times$ is the adjoint with respect to the real Krein-Schmidt product. Note that the adjoint of a $\CC$-linear operator with respect to the real and complex Krein-Schmidt products coincide, so that this notation is consistent.} $P=P^\times=P^2$. Moreover, for all $a\in \A_F$, $v\in V$ and $w\in V^\perp$ we have
\bea
(v,\pi_F(a)w)_\RR&=&(\pi_F(a)^\times v, w)_\RR,\mbox{ by property of the Krein-Schmidt product}\cr
&=&(\pi_F(a^*)v,w)_\RR,\mbox{ since }\B_F\mbox { is a }*-\mbox{background}\cr
&=&0,\mbox{ since }V\mbox { is an }\A_F-\mbox{bimodule}
\eea
Thus $V^\perp$ is a left $\A_F$-module, and we can similarly prove that it is a right $\A_F$-module. It follows that for any $a,b\in \pi_F(\A_F)$ and $T\in\End(S\otimes \K_F)$, we have
\be 
P(aTb)=aP(T)b\label{Pmorph}
\ee
Now, let $\omega\in \Omega^1$. The curvature $\rho_D(\omega)$ is a function on $M$ with values in $\End(S\otimes \K_F)/V$. Thus for all $x\in M$, $P(\rho_D(\omega)_x)$ is a well-defined element of $V^\perp\subset \End(S\otimes \K_F)$, and the \emph{generalized Connes-Lott bosonic action} 
\begin{equation}\label{ba1}
S_b(\omega):=-(P(\rho_D(\omega)),P(\rho_D(\omega)))=-\int_M\tr(P(\rho_D(\omega)_x)^\times P(\rho_D(\omega)_x)){\rm vol}_g
\end{equation}
is a well-defined function on $\Omega^1$, which is invariant under gauge-transformations \eqref{gaugetrans} by virtue of \eqref{gaugecov} and \eqref{Pmorph}.

Let us consider an origin Dirac operator $D$ and the following subspaces of the configuration space:
\bea
\D_{\hat\Omega^1}&:=&\hat\Omega^1\cap \D_\B\cr
\D_{\Omega^1}&:=&\{\hat\omega\in \D_{\hat\Omega^1}|\exists \omega\in \Omega^1, \hat\omega=\omega+\omega^o\}\nonumber
\eea
Dirac operators of the form $D+\hat\omega$ with $\hat\omega\in \D_{\Omega^1}$ (resp. with $\hat\omega\in \D_{\hat\Omega^1}$) are called \emph{fluctuations} (resp.  \emph{generalized fluctuations}) of $D$. It is easy to see that the affine space of generalized fluctuations is invariant under $\G_\A$, since $\Upsilon(u)D\Upsilon(u)^{-1}-D=\Upsilon(u)[D,\Upsilon(u^{-1})]\in \D_{\hat\Omega^1}$ for any $u\in U(\A)$. The action of $\Ad_{\Upsilon(u)}$ on the element $D+\hat\omega$ of $\D_{\hat \Omega^1}$ is the gauge-transformation $\hat\omega\rightarrow \hat\omega^{\Upsilon(u)}$ given by formula \eqref{gaugetrans}, and the discussion above applied to the $J$-symmetrized background $\hat \B$ shows that the Connes-Lott action is invariant under these transformations. We thus see that there is a well-defined theory, with configuration space $D+\D_{\hat\Omega^1}$ and action \eqref{ba1}, under the hypotheses $C_0$ (which have used all along) and weak $C_1$, which ensures that $\G_\A\subset\Aut(\B)$. If the order $1$ condition holds, then the affine space of fluctuations is also invariant under $\G_\A$, and thanks to $C_1$ we have $(\omega+\omega^o)^{\Upsilon(u)}=\omega^u+(\omega^u)^o$.


\section{The   configuration space of the NCSM}\label{configspace}
The total configuration space  of the algebraic background $\B$ of the Lorentzian NCSM has been computed in \cite{algbackpart2}. It is of the form
\be 
\D_\B=\D_{\rm Gravity}\oplus\D_{\Omega^1}\oplus  \D_{B-L}\oplus\D_\sigma\oplus \D_{\rm flavour}.\label{decompconfig}
\ee

The gravity part contains Dirac operators of the form $\delta_e\hat\otimes 1$, where $\delta_e$ is the (rescaled, see \cite{algbackpart1}) canonical Dirac operator associated with a  tetrad $e$.

As for $\D_{\Omega^1}$, it can be decomposed according to \eqref{decomp1form} into $\D_{\rm  Gauge  }\oplus \D_{\rm Higgs}$.  The gauge part contains the gauge fields, that is, gauge fields of the Standard Model, plus an anomalous field coming from the extra $U(1)$ part of the unitary group of $\A_F$, which we write $U(1)_X$ in what follows. The elements of $\D_{\rm Gauge}$ are of the form
\be 
i\gamma^\mu\hat\otimes A_\mu+(i\gamma^\mu\hat\otimes A_\mu)^o=i\gamma^\mu\hat\otimes (A_\mu-A_\mu^o)=i\gamma^\mu\hat\otimes (A_\mu+J_FA_\mu J_F^{-1})
\ee
where $A_\mu$ is a  field with values in $\pi_F(\A_F)$ which is Krein anti-selfadjoint. Hence $A_\mu+J_FA_\mu J_F^{-1}$ is in the Lie algebra of $\G_{\A_F}$, which has the following basis:
\bea
t_X&=&[\colvec{0&0\cr 0&-2i}\oplus \colvec{0&0\cr 0&-2i}\otimes 1_3,-i1_2\oplus -i1_2\otimes 1_3,c,c]\otimes 1_N\cr
t_Y&=&[\colvec{0&0\cr 0&-2i}\oplus \colvec{{4i\over 3}&0\cr 0&-{2i\over 3}}\otimes 1_3,-i1_2\oplus{i\over 3}1_2\otimes 1_3,c,c]\otimes 1_N\cr
t_{W}^a&=&[0,i\sigma^a\oplus i\sigma^a\otimes 1_3,c,c]\otimes 1_N, a=1,2,3\cr
t_{C}^a&=&[0\oplus 1_2\otimes i {\lambda^a},0\oplus 1_2\otimes i\lambda^a,c,c]\otimes 1_N, a=1,\ldots,8\label{othergen}
\eea
where $c,c$ stand for the complex conjugates of the two first entries (hence all matrices have the form $[a,b,a^*,b^*]$), and where we may choose the bases $\sigma^a$ and $\lambda^a$ of Pauli and Gell-Mann matrices, respectively. Moreover, we can show by direct inspection that $(t_Y,t_W^a,t_C^b)$ is an orthogonal basis with respect to the Krein-Schmidt product, which  restricts to an invariant scalar product on the Lie algebra of $\G_{\A_F}$. Let us now introduce some notations. We write:
\be 
A_\mu+J_FA_\mu J_F^{-1}=\BB_\mu=\BB_\mu^Yt_Y+\BB^W_{\mu a}t_W^a+\BB^C_{\mu a}t_C^a\label{defbmu}
\ee
\begin{rem}
Note that the field $\BB_\mu$ is Lie algebra-valued and thus anti-selfadjoint, whereas the fields used by physicists are selfadjoint and defined by $\AA_\mu=i\BB_\mu$. The definition of curvature must change accordingly. For $\BB_\mu$ it is 
\be
{\mathbb F}_{\mu\nu}=\partial_\mu {\mathbb B}_\nu-\partial_\nu {\mathbb B}_\mu+[{\mathbb B}_\mu,{\mathbb B}_\nu]\label{defcurv}
\ee
so that $\FF_{\mu\nu}=-i(\partial_\mu\AA_\nu-\partial_\nu \AA_\mu-i[\AA_\mu,\AA_\nu])$, and the definition of the curvature of $\AA$ used by physicists is   the expression between parentheses. This change of convention from what is usual in physics is imposed on us by  the factor of $i$ which is included in the elements of $\Omega^1_M$.
\end{rem}

Let us now consider the elements of $\D_{\rm Higgs}$. They can be written $1\hat\otimes(\Phi(q)+\Phi(q)^0)$ where
\be
\Phi(q)=\begin{pmatrix}0&-Y_0^\dagger\tilde q^\dagger&0&0\cr \tilde q Y_0&0&0&0\cr 0&0&0&0\cr 0&0&0&0
\end{pmatrix}
\ee
is a generic Krein selfadjoint finite $1$-form, and $q$ is a $\HH$-valued field.   

Let us turn to the elements of $\D_{B-L}$. They are of the form
\be 
i\gamma(v)\hat\otimes t_{B-L},
\ee
where $v$ is a vector field and the $t_{B-L}$ is the anti-selfadjoint generator of $B-L$ symmetry, that is
\be 
t_{B-L}=i[-1_2\otimes{1\over 3}1_2\otimes 1_3,-1_2\otimes{1\over 3}1_2\otimes 1_3,1_2\otimes{-1\over 3}1_2\otimes 1_3,1_2\otimes{-1\over 3}1_2\otimes 1_3]\otimes 1_N\label{tbl}
\ee
To describe the elements of $\D_\sigma$, first define $\sigma(m)$ for $m\in M_N(\CC)$, by
\be 
\sigma(m)=\begin{pmatrix}
0&0&sp_\nu\otimes m^\dagger&0\cr
0&0&0&0\cr
p_v\otimes m&0&0&0\cr
0&0&0&0
\end{pmatrix}\label{defsigma}
\ee
Then the elements of $\D_\sigma$ are of the form $1\hat\otimes\sigma(m)$ where $m$ is smooth field of matrices satisfying $m^T=-sm$.

Finally we come to $\D_{\rm flavour}$. It contains multivector  flavour changing fields which we do not need to describe precisely here.  For more details see \cite{algbackpart2}.

Now let us consider the effect of the symmetries on the decomposition \eqref{decompconfig}. First, diffeomorphisms and spinomorphisms preserve all the summands. Local gauge symmetries stabilize $\D_{\rm Gauge}$ and $\D_{\rm Higgs}$ and commute with elements of $D_{B-L},\D_\sigma$ and $\D_{\rm flavour}$. Moreover, for a given tetrad $e$, we have, using $C_0$ and $C_1$:
\be
\Upsilon(u)\delta_e\hat\otimes 1 \Upsilon(u)^{-1}=\delta_e\hat\otimes 1+\omega+\omega^o
\ee
with $\omega=\pi(u)[\delta_e\hat\otimes 1,\pi(u^{-1})]\in \Omega^1$. Thus $\Upsilon(u)\delta_e\hat\otimes 1 \Upsilon(u)^{-1}\subset\delta_e\hat\otimes 1+\D_{\rm Gauge}$.

Local $B-L$ symmetries commute with the elements of $\D_{\rm Gauge}$, $\D_{B-L}$, $\D_{\rm Higgs}$,  $\D_{\rm flavour}$, and with the  notations of section 2, 
\bea
U_\varphi\delta_e\hat\otimes 1 U_\varphi^{-1}&=&\delta_e\hat\otimes 1 -i\gamma(\nabla\varphi)\hat\otimes t^1_{B-L}\cr
&\subset&\delta_e\hat\otimes 1+\D_{B-L}\label{extgau1}
\eea
Their action on $1\hat\otimes\sigma(m)\in \D_\sigma$ is 
\be 
U_\varphi  1\hat\otimes\sigma(m) U_\varphi^{-1}=1\hat\otimes \sigma(e^{2i\varphi}m)\label{extgau2}
\ee
We are now going to define several restricted configuration spaces which will be of interest in this paper. First,  since we are not concerned here with the dynamics of the gravitational field, we will consider only a single tetrad $e$ and the corresponding element $D_M\hat\otimes 1:=\delta_e\hat\otimes 1\in \D_{\rm Gravity}$. This forces us to restrict the symmetry group by suppressing spinomorphisms and allowing only the diffeomorphisms which preserve the metric $g_e$ defined by $e$. Next we set all flavour fields to $0$: we see that it is allowed by the symmetries. We also consider at most one complex $\sigma$-scalar field in this paper, which we can do thanks to \eqref{extgau2}. We do this out of simplicity, and also because only one such field appears in the extended model we will define below. Hence we fix a non-zero matrix in $M_N(\CC)$, which for consistency must be   $m_0$, and consider the $2$-dimensional real subspace $\D_{m_0}^\CC$ of $\D_\sigma$ containing only the elements which are of the form $1\hat\otimes\sigma(zm_0)$, $z\in\CC$.  In the sequel we write them simply $1\hat\otimes \sigma(z)$, $m_0$ being understood. We have thus restricted the configuration space to 
\be 
\D_{\rm SM+X}^{\rm ext}:=D_M\hat\otimes 1\oplus\D_{\rm Gauge}\oplus \D_{B-L}\oplus \D_{\rm Higgs}\oplus\D^\CC_{m_0}\label{DSMextplusX}
\ee
and the symmetry group to global isometries, local gauge and $B-L$ transformations. However,  $\D_{\rm Gauge}$ contains an anomalous part coming from $t_X$. Hence we enforce the \emph{unimodularity condition}, which amounts to set the $X$-field to zero by hand. We are then forced to restrict the local gauge symmetries to   unimodular ones, that is, those of the form $\Upsilon(u)$ with $\det\pi_F(u(x))=1$ for all $x\in M$. Doing so we obtain  the $19$-dimensional real affine space
\be 
\D_{\rm SM}^{\rm ext}:=D_M\hat\otimes 1\oplus\D_{\rm Gauge}^{\rm Tracefree}\oplus \D_{B-L}\oplus \D_{\rm Higgs}\oplus\D^\CC_{m_0}.\label{DSMext}
\ee
This is the configuration space of the extended NCSM which is the main subject of this paper. The bosonic fields it contains are the gauge bosons of the Standard Model plus a $Z_{B-L}'$-boson and a complex scalar field of charge $2$ under $B-L$. It is interesting to observe that the exact same field content has been derived from a non-standard version of NCG in \cite{BF2}. The corresponding model had been previously studied in \cite{boylecosmo}.  

The NCSM itself has the bosonic configuration space 
\be 
\D_{\rm SM} :=(D_M\hat\otimes 1+1\hat\otimes\sigma(1))\oplus\D_{\rm Gauge}^{\rm Tracefree}\oplus  \D_{\rm Higgs} 
\ee
It is only stable under the group   generated by global $g_e$-isometries and local unimodular gauge symmetries.  Though it is not very natural from the point of view of algebraic backgrounds, we will consider it as a benchmark.

\section{The bosonic action of the Lorentzian NCSM}
The elements of $\D_{\rm SM}$ can be written in the form
\bea
&&(D_M\hat\otimes 1+1\hat\otimes\sigma(1))+i\gamma^\mu\hat\otimes \BB_\mu+1\hat\otimes(\Phi(q)+\Phi(q)^o)\cr
&=&(D_M\hat\otimes 1+1\hat\otimes D_F)+i\gamma^\mu\hat\otimes \BB_\mu+1\hat\otimes(\Phi(q-1)+\Phi(q-1)^o)\cr
&=&D+\omega+\omega^o\hspace{3cm} 
\eea
where $\omega$ is a Krein selfadjoint element of $\Omega^1$. Thus, the bosonic fields can be described as fluctuations of  $D=D_M\hat\otimes 1+1\hat\otimes D_F$. Moreover it can be checked that $\Omega^1\cap (\Omega^1)^o=\{0\}$, so that the bosonic degrees of freedom can be parametrized by a Krein selfadjoint $1$-form $\omega$. Hence, to define the bosonic action of the Lorentzian NCSM we \emph{could} use 
\begin{equation}
S_b(\omega)=-(P(\rho_D(\omega)),P(\rho_D(\omega)))\label{baCL}
\end{equation}
just as in the Euclidean Connes-Lott theory.  However it should be clear from the above discussion that the true bosonic variable  is $\hat\omega:=\omega+\omega^o$. Moreover this is precisely $\hat\omega$ which couples to fermions, and anyway in the B-L theory we will no longer have the choice since $\Omega^1_{\rm ext}\cap (\Omega^1_{\rm ext})^o\not=\{0\}$. We thus prefer to use $\hat\omega$ directly. This means that we need to compute its curvature, which is meaningful in the $J$-symmetrized algebraic background $\hat\B$. We thus define the bosonic action of the NCSM to be the generalized Connes-Lott action given by formula \eqref{ba1} which we repeat here:
\be 
S_b(\hat\omega)=-(P(\rho_D(\hat\omega)),P(\rho_D(\hat\omega)))\label{baNCSM}
\ee
Despite the apparent similarity between \eqref{baCL} and \eqref{baNCSM}, they involve quite different calculations since both the junk and the algebra are different. It is thus   remarkable that these two expressions actually agree up to an overall factor of $2$, and a numerical factor in front of the Higgs term (the reason is essentially lemma \ref{curvomegaz}). It yields the exact terms of the SM bosonic action, with the correct signs. This will be clear from the calculations of section \ref{extba}, which are more general, but can be checked independently as an exercise.

%
%
%
%

\section{The extended algebraic background}\label{secextab}
As previously remarked, $\D_{\rm SM}$ is not very natural and  we would prefer to use $\D_{\rm SM}^\ext$, but since its elements are not fluctuations or even generalized fluctuations, we cannot use Connes-Lott action on it. However, there is a simple extension $\B_F^\ext$ of the finite background which is such that the total background $\B^\ext=\B_M\hat\otimes\B_F^\ext$ has the following remarkable properties: its configuration space and automorphism group are the same as that of the NCSM, except that local $B-L$ symmetries belong to the gauge group and $\D_{\rm SM}^\ext$ only contains fluctuations of $D$, so that one can use the Connes-Lott action.

The finite background $\B_F^\ext$ only differs from $\B_F$ by the algebra, representation, and bimodule of $1$-forms. Let us first define the \emph{extended finite algebra}  $\A_F^\ext:=\A_F\oplus \CC$. It is represented on $\K_F$ by
\begin{equation}
\pi_F^\ext(\lambda,q,m,\mu)=[\tilde q_\lambda,\tilde q,\mu\otimes 1_N\oplus 1_2\otimes m\otimes 1_N,\mu\otimes 1_3\oplus 1_2\otimes m\otimes 1_N]\label{generalext}
\end{equation}
The bimodule of extended $1$-forms $(\Omega^1_F)^\ext$ is defined by the requirement that $D_F$ still be a regular Dirac. Hence  $(\Omega^1_F)^\ext$ is the $\pi_F^\ext(\A_F^\ext)$-bimodule generated by $[D_F,\pi_F^\ext(b)]$, $b\in\A_F^\ext$. Let us   describe more precisely the extended $1$-forms. Since the projector on   anti-lepton space   $p_{\bar\ell}$ is equal to $\pi_F^\ext(0,0,0,1)$, we have $\pi_F^\ext(\A_F^\ext)=\pi_F(\A_F)\oplus \CC p_{\bar\ell}$. Moreover $p_{\bar\ell}$ commutes with $\pi_F(\A_F)$ and $p_{\bar\ell}\omega=\omega p_{\bar\ell}=0$ for any $\omega\in\Omega^1_{F}$.  An extended finite 1-form is thus  a sum of terms of the form:
\bea
a[D_F,b]+zp_{\bar\ell}[D_F,c]+z'd[D_F,p_{\bar \ell}]+z''p_{\bar\ell}[D_F,p_{\bar \ell}]&:=&\cr
\omega_1+z'd\colvec{0&0&2sM_0^\dagger&0\cr 0&0&0&0\cr -2M_0&0&0&0\cr 0&0&0&0}+z''p_{\bar\ell}\colvec{0&0&0&0\cr 0&0&0&0\cr -2M_0&0&0&0\cr 0&0&0&0}
&=&\cr
\omega_1+\colvec{0&0&2z'\lambda sM_0^\dagger&0\cr 0&0&0&0\cr -2z'\lambda M_0&0&0&0\cr 0&0&0&0}+ \colvec{0&0&0&0\cr 0&0&0&0\cr -2z''M_0&0&0&0\cr 0&0&0&0}
\eea
where  $a,b,c,d\in \pi_F(\A_F),\lambda, z,z',z''\in\CC, \omega_1\in\Omega^1_{F}$. Hence we conclude that :
\begin{propo}
We have $(\Omega^1_F)^\ext=\Omega^1_F\oplus \Omega^1_\sigma$, where   $\Omega^1_\sigma\simeq \CC\oplus\CC$ is the bimodule of elements of the form 
\be 
\colvec{0&0&z_2M_0^\dagger&0\cr 0&0&0&0\cr z_1M_0&0&0&0\cr 0&0&0&0}\label{genomegasigma}
\ee
with $z_1,z_2\in\CC$.
\end{propo}
Observe that  $(\Omega^1_\sigma)^o=\Omega^1_\sigma$, so that $\Omega^1_\sigma$ is both a $\pi_F^\ext(\A_F^\ext)$ and a $\pi_F^\ext(\A_F^\ext)^o$-bimodule. Note also that the $1$-form \eqref{genomegasigma} is Krein-self adjoint iff $z_2=s\bar z_1$, hence is of the form $\sigma(z)$, $z\in\CC$ (see \eqref{defsigma}). Another interesting property is $\sigma(z)^o=\sigma(z)$. Let us now show that $\B_F^\ext$ satisfies $C_0$ and weak $C_1$. A general element of $\pi_F^\ext(\A_F^\ext)^o$ is of the form 
\be
[\mu'\otimes 1_N\oplus 1_2\otimes m'\otimes 1_N,\mu'\otimes 1_N\oplus 1_2\otimes m'\otimes 1_N,\tilde q_{\lambda'},\tilde q']
\ee
which clearly commutes with \eqref{generalext}.  Such an element also commutes with $\Omega^1_F$, but its commutator with \eqref{genomegasigma} is found to be $\colvec{0&0&(\mu'-\lambda')z_2M_0^\dagger&0\cr 0&0&0&0\cr (\lambda'-\mu')z_1M_0&0&0&0\cr 0&0&0&0}$.  Hence $\B_F^\ext$ does not satisfy the order $1$ condition. Instead we have the relation
\be 
[\pi_F^\ext(\A_F^\ext)^o,(\Omega^1_F)^\ext]=[p_{\bar \ell}^o,\Omega^1_\sigma]=\Omega^1_\sigma,\label{rel}
\ee
from which the weak $C_1$ condition immediately follows.

Let us now compute   the \emph{extended gauge group}, which we write $\G_{\A_F}^\ext$ instead of $\G_{\A_F^\ext}$, for ease of reading. Let $u=(e^{i\theta},q,m,e^{i\varphi})\in U(\A_F^\ext)$, with $q\in SU(2)$ and $m\in U(3)$. In order  to correctly identify the  $U(1)$ of weak hypercharge, we write $\varphi=\theta+t$. We will also need to decompose (non uniquely) $m$ as $m=e^{i\xi}g$, with $g\in SU(3)$ and $\xi\in \RR$. With these notations we have $\Upsilon(u)=[A,B,A^*, B^*]\otimes 1_N$, where
\bea
A&=&q_{e^{i\theta}}e^{-i(t+\theta)}\oplus q_{e^{i\theta}}e^{-i\xi}\otimes \bar g =\colvec{e^{-it}&0\cr 0&e^{-i(t+2\theta)}}\oplus \colvec{e^{i(\theta-\xi)}&0\cr 0&e^{-i(\theta+\xi)}}\otimes \bar g\cr
B&=&qe^{-i(t+\theta)}\oplus qe^{-i\xi}\otimes \bar g\label{genelextgg}
\eea
Writing $f(e^{i\theta},q,g,e^{it},e^{i\xi})=\Upsilon(u)$, we see that $f$ is a surjective homomorphism from $U(1)\times SU(2)\times SU(3)\times U(1)\times U(1)$ to $\G_{\A_F}^\ext$. One readily computes that $\det(\pi_F^\ext(u))=e^{4Ni(t+\theta+3\xi)}$. After the choice of a cubic root of unity, we can write 
\bea
f(e^{i\theta},q,g,e^{it},e^{i\xi})&=&f(e^{i\theta},q,g,e^{it},e^{-i{\theta+t\over 3}})f(1,1,1,1,e^{i(\xi+{\theta+t\over 3})})
\eea
where the first factor belongs to the unimodular extended gauge group $S\G_{\A_F}^\ext$ and can be further decomposed into
\be 
f(e^{i\theta},q,g,e^{it},e^{-i{\theta+t\over 3}})=f(e^{i\theta},q,g,1,e^{-i{\theta\over 3}})f(1,1,1,e^{it},e^{-i{t\over 3}})
\ee
Using  \eqref{genunimod}, we see that it is the product of an element of $S\G_{\A_F}$ with  $g_{B-L}(t)$. Hence,  the extended gauge group is generated by the subgroups $S\G_{\A_F}$, $U(1)_{B-L}$ and $U(1)_X$, which commute with one another. However,  there is an ambiguity in the decomposition of an element of $\G_{\A_F}^\ext$ into the product of  factors belonging to these subgroups, which corresponds to taking the quotient by a finite abelian group. We do not need more precision since we will soon deal with the Lie algebra.

\begin{rem} The abelian factor of the extended gauge group can be identified with $U(1)^3$ (modulo the quotient by a finite group), in an infinite number of ways. However, the unimodularity condition singles out the decomposition $U(1)_X\times (U(1)_{Y}\times U(1)_{B-L})$.
\end{rem}

Observe that $\G_{\A_F}^\ext=\Aut(\B_F)$. But what is $\Aut(\B_F^\ext)$ ? An element $U\in\Aut(\B_F^\ext)$ induces an automorphism of $\B_F^\ext$, which must be inner. There thus exists a  $g\in \G_{\A_F}^\ext$ such that $U'=Ug^{-1}$ commutes with $\pi_F^\ext(\A_F^\ext)$. By proposition 15 in \cite{algbackpart2}, $U'$ is a flavour symmetry, i.e. $U'=[A,B,A^*,B^*]$, with 
\bea
A&=&(p_\nu\otimes g_\nu+ p_e\otimes g_e)\oplus( p_u\otimes 1_3\otimes g_u+p_d\otimes 1_3\otimes g_d)\cr
B&=&1_2\otimes g_\ell\oplus 1_2\otimes 1_3\otimes g_q
\eea
where $g_\nu,\ldots,g_q$ are unitary matrices. Now $\Ad_{U'}$ must preserve $(\Omega^1)^\ext$, and it is immediate that it must preserve $\Omega^1_F$ and $\Omega^1_\sigma$ separately. From proposition 17 in \cite{algbackpart2}, we obtain from this that $U'\in \G_{\A_F}^\ext$, so that  $\Aut(\B_F^\ext)= \G_{\A_F}^\ext=\Aut(\B_F)$.

We thus see that the extension of the finite background unifies the symmetries, which are now all of gauge type. It can also be shown, by copying \emph{verbatim} the calculation of $\Aut(\B)$ in \cite{algbackpart2}, section 6.5, that $\Aut(\B^\ext)=\Aut(\B)$.
 
Let us now turn to the extended gauge fields. They are by definition of the form $\omega+\omega^o$ where $\omega$ is a selfadjoint extended 1-form in the manifold part of $\Omega^1_\ext$, that is $\Omega^1_M\hat\otimes \pi_F^\ext(\A_F^\ext)$. For the same reason as for the Standard Model, it means that they are of the form 
\be 
i\gamma^\mu\hat\otimes B_{\mu,a} t^a\label{genfield}
\ee
where $t^a$ runs through a basis of the Lie algebra of the extended gauge group, and $B_{\mu,a}$ are real fields. The basis $t^a$ we will use (in order to recover the usual fields) is $t_X,t_Y,t^a_W,t^a_C,t_{B-L}$ already defined in equations \eqref{othergen} and \eqref{tbl}. The Krein-Schmidt product restricts to an  invariant scalar product on this Lie algebra. The scalar products among basis elements will be important later. As already noted, $t_Y,t^a_W$ and $t^a_C$ are orthogonal to each other, and it is immediate that $t_{B-L}$ is orthogonal to $t^a_W$ and $t^a_C$. However $t_Y$ and $t_{B-L}$ are not orthogonal. Their scalar product is
\bea
\tr(t_Y^\times t_{B-L})&=&-\tr(t_Yt_{B-L})\cr
&=&-2 \tr(\colvec{0&0\cr 0&-2}+\colvec{-4/9&0\cr 0&2/9}\otimes 1_3+(-1)1_2+(-{1\over 9})1_2\otimes 1_3)N\cr
&=&{32N\over 3}
\eea
Other useful traces are (using $\tr(\lambda^a)^2=2$):
\bea
\tr(t_Y^\times t_Y)&=&{80\over 3}N\cr
\tr((t_W^a)^\times t_W^a)&=&16N\cr
\tr((t_C^a)^\times t_C^a)&=&16N\cr
\tr(t_{B-L}^\times t_{B-L})&=&{32\over 3}N
\eea
We will write $Z'$ for the component along $t_{B-L}$ with respect to the basis $t^a$, so that an unimodular extended gauge field of the form \eqref{genfield} can be written:
\be 
i\gamma^\mu\hat\otimes(\BB_\mu+Z_\mu't_{B-L})=i\gamma^\mu\hat\otimes(\BB_\mu^Yt_Y+\BB^W_{\mu a}t_W^a+\BB^C_{\mu a}t_C^a+Z_\mu't_{B-L})
\ee
When $\omega$ is in the finite part of $\Omega^1_\ext$,  the fluctuation $\omega+\omega^o$ will contribute by scalar fields of the form $1\hat\otimes \Phi(q')+1\hat\otimes \Phi(q')^o+1\hat\otimes \sigma(z')$, where $q'$ and $z'$ are quaternion and complex fields respectively. Hence, a general fluctuation of the Dirac operator $D=D_M\hat\otimes 1+1\hat\otimes D_F$ will be
\bea
D+\omega+\omega^o&=&D+i\gamma^\mu\hat\otimes (\BB_\mu+Z_{\mu}'t_{B-L})+1\hat\otimes \Phi(q')+1\hat\otimes \Phi(q')^o+1\hat\otimes \sigma(z')\cr
&=&D_M\hat\otimes 1+i\gamma^\mu\hat\otimes  (\BB_\mu+Z_{\mu}'t_{B-L})+1\hat\otimes \Phi(q)+1\hat\otimes \Phi(q)^o+1\hat\otimes \sigma(z)\cr
&&\label{extflu}
\eea
with $q'=q-1$ and $z'=z-1$, and where $\omega$ is a general selfadjoint element of $\Omega^1_\ext$. We thus see that $\D_{(\Omega^1)^\ext}=\D_{\rm SM+X}^\ext$. Since $\B^\ext$ does not satisfy $C_1$, this space is not guaranteed to be stable under local extended gauge transformations. However, we know it is thanks to \eqref{extgau1} and \eqref{extgau2}. Note that  a local $B-L$ transformation $1\hat\otimes e^{\varphi t_{B-L}}$ will boil down to 
\bea
Z_\mu'&\rightarrow&Z_\mu'-\partial_\mu\varphi\cr
z&\rightarrow&e^{2i\varphi}z
\eea
so that $z'$ goes to $e^{2i\varphi}(z'+1)-1$. As usual we will remove the $X$-field by the unimodularity condition, and we will obtain a configuration space ($\D_{\rm SM}^\ext$) and symmetry group (the unimodular extended gauge transformations) such that the generalized Connes-Lott action is well-defined and invariant.

\begin{rem}[1]
We see that the success of the approach is kind of accidental. However it can be put in another perspective. Since  $\B^\ext$ satisfies weak $C_1$, we know  from section \ref{curvform} that  the generalized Connes-Lott action is well-defined and gauge-invariant on $\D_{(\hat\Omega^1)^\ext}$. However, it can be shown that $\D_{(\hat\Omega^1)^\ext}=\D_{(\Omega^1)^\ext}\oplus Z$, where $Z$ is a 3 dimensional space of abelian gauge fields, which are all gauge-invariant and do not appear under gauge transformations. It thus follows that $\D_{(\Omega^1)^\ext}$ itself is gauge-invariant.
\end{rem}
%


\begin{rem}[2] In \cite{pertsemigroup}, a theory of inner fluctuations in the absence of the order 1 condition (the perturbation semi-group) has been developped. In brief, in such a context one ought to supplement the usual terms $\omega+\omega^o$, where $\omega$ is a self-adjoint 1-form, with an additional term of the form
\be
\omega_s=\sum_{j} J a_j[\omega^o, b_j]J^{-1}\label{pert}
\ee
where $\omega=\sum_j a_j[D,b_j]$. There is thus a non-linear map from the $1$-forms to the configuration space, and  it can be shown that in the present context  the linearity boils down to the replacement of $z$ by $z^2$. In the Euclidean case, the spectral action can then be computed on $D+\omega+\omega^o+\omega_s$. One would obtain a submodel of the noncommutative Pati-Salam theory considered in \cite{ccvs} and it would be interesting to compare the results with the ones we obtain below.  
\end{rem}

\section{The bosonic action}\label{extba}
\subsection{The extended $J$-symmetrized algebras and forms}
In this section we just quote results which are proved in appendix \ref{appA}. The $J$-symmetrized algebras $\hat\A_F$ and $\hat \A_F^\ext$ contain, respectively, the elements of the form
\be
a=[\colvec{z_\nu &0\cr 0&z_e}\oplus \colvec{m_u&0\cr 0&m_d},\alpha\oplus\beta,\colvec{z_\nu & 0\cr 0& z_e^*}\oplus \colvec{m_{\bar u}&0\cr 0&m_{\bar d}},\gamma\oplus\delta]\otimes 1_N
\ee
where $z_\nu,z_e\in\CC$, $m_u,\ldots, m_{\bar d}\in M_3(\CC)$, $\alpha,\gamma\in M_2(\CC)$ and $\beta,\delta\in M_2(\CC)\otimes M_3(\CC)$, and 
\be
b=[\colvec{z_\nu &0\cr 0&z_e}\oplus \colvec{m_u&0\cr 0&m_d},\alpha\oplus\beta,\colvec{z_{\bar \nu} & 0\cr 0&  z_{\bar e}}\oplus \colvec{m_{\bar u}&0\cr 0&m_{\bar d}},\gamma\oplus\delta]\otimes 1_N
\ee
where now $z_\nu,z_e,z_{\bar \nu}, z_{\bar e}$ are four independent complex numbers. 
The  elements  of $\hat\J^1_{D_F}$ are of the form
\be
\Phi=[0, \alpha'\otimes(Y_\nu Y_\nu^\dagger-Y_e Y_e^\dagger)\oplus \beta'\otimes (Y_u Y_u^\dagger-Y_d Y_d^\dagger),0,\gamma'\otimes( Y_\nu^* Y_\nu^T- Y_e^* Y_e^T) \oplus \delta'\otimes (Y_u^* Y_u^T- Y_d^* Y_d^T) ]
\ee
with $\alpha',\gamma'\in M_2(\CC)$, and $\beta',\delta'\in  M_2(\CC)\otimes M_3(\CC)$. An element of the total junk $\hat \J^1_D$  is a function with values in $\hat\A_F+\J^1_{D_F}$, that is, with values of the form $[A,B,C,D]$ with
\begin{align*}
A&=\colvec{z_\nu &0\cr 0&z_e}\otimes 1_N\oplus \colvec{m_u&0\cr 0&m_d}\otimes 1_N,\cr
B&=(\alpha\otimes 1_N+\alpha'\otimes(Y_\nu Y_\nu^\dagger-Y_e Y_e^\dagger))\oplus(\beta\otimes 1_N+\beta'\otimes (Y_u Y_u^\dagger-Y_d Y_d^\dagger)),\cr
C&=\colvec{z_\nu & 0\cr 0&z_e^*}\otimes 1_N\oplus \colvec{m_{\bar u}&0\cr 0&m_{\bar d}}\otimes 1_N,\cr
D&=(\gamma\otimes 1_N+\gamma'\otimes( Y_\nu^* Y_\nu^T- Y_e^* Y_e^T))\oplus(\delta\otimes 1_N+\delta'\otimes ( Y_u^* Y_u^T- Y_d^* Y_d^T)).
\end{align*}
Let us first remark that since these elements are diagonal, they commute with the fundamental symmetry, hence the Krein-Schmidt product is positive definite on them. We can also see that the genereticity hypothesis ensures that the sum $\hat \A_F+\hat\J^1_{D_F}$ is direct, but it is orthogonal only if $\tr Y_e Y_e^\dagger=\tr Y_\nu Y_\nu^\dagger$ and $\tr Y_u Y_u^\dagger=\tr Y_d Y_d^\dagger$. In this  special case, the projection on the orthogonal of the junk will have a different form. \emph{In the sequel we suppose that $\tr Y_e Y_e^\dagger\not=\tr Y_\nu Y_\nu^\dagger$ and $\tr Y_u Y_u^\dagger\not=\tr Y_d Y_d^\dagger$.} Let us also introduce the following useful notation: for any matrix $A\in M_N(\CC)$, we write:
\be
\tilde A=A-{\tr(A)\over N}1_N
\ee
for the traceless part of $A$.

Let us turn to $(\hat\J^1_{D_F})^\ext$. One can show that $(\hat\J^1_{D_F})^\ext=\hat\J^1_{\rm diag}\oplus \M$, where the sum is orthogonal, $\hat\J^1_{\rm diag}$ contains the elements of the form $[A,B,C,D]$ with
\begin{align*}
A&=\colvec{z_\nu &0\cr 0&z_e}\otimes 1_N\oplus \colvec{m_u&0\cr 0&m_d}\otimes 1_N,\cr
B&=(\alpha\otimes 1_N+\alpha'\otimes(Y_\nu Y_\nu^\dagger-Y_e Y_e^\dagger))\oplus(\beta\otimes 1_N+\beta'\otimes (Y_u Y_u^\dagger-Y_d Y_d^\dagger)),\cr
C&=\colvec{z_{\bar \nu} & 0\cr 0&z_{\bar e}}\otimes 1_N\oplus \colvec{m_{\bar u}&0\cr 0&m_{\bar d}}\otimes 1_N,\cr
D&=(\gamma\otimes 1_N+\gamma'\otimes( Y_\nu^* Y_\nu^T- Y_e^* Y_e^T))\oplus(\delta\otimes 1_N+\delta'\otimes ( Y_u^* Y_u^T- Y_d^* Y_d^T)),
\end{align*}
and $\M\simeq M_2(\CC)^4$ is the module of antidiagonal elements of the form
\be
\rho= \colvec{0&0&0&M_0^\dagger Y_\ell^T\gamma_1\cr 0&0& \alpha_1Y_\ell M_0^\dagger&0\cr 0& M_0Y_\ell^\dagger \gamma_2&0&0\cr \alpha_2  Y_\ell^* M_0&0&0&0}
\ee
with $\alpha_1,\ldots,\gamma_2\in M_2(\CC)$.

\subsection{The curvature 2-form}
Let us write the field $\hat \omega$  in the form

\bea
\hat\omega&=&\omega_{SM}+\omega_{SM}^o+\omega_\sigma
\eea
where 
\bea
\omega_{SM}&=&i\gamma^\mu\hat\otimes A_\mu+1\hat\otimes \Phi(q'),\mbox{ with }q'=q-1\cr
\omega_\sigma&=&i\gamma^\mu\hat\otimes Z'_\mu t_{B-L} +1\hat\otimes \sigma(z')\mbox{ with }z'=z-1
\eea
using the notations of section \ref{configspace}. Note that here $\tr(A_\mu)=0$ so that $\BB_\mu=A_\mu-A_\mu^o$ is an unimodular gauge field. Let us now prove a useful lemma.

\begin{lemma}\label{lemsympa} Let $A_1,A_2$ be two commuting   subalgebras of $A_3$. Suppose that for all $a_1\in A_1$ and $a_2\in A_2$ one has $a_1[D,a_2]=[D,a_2]a_1$. Then the 1-forms of $A_1$ and $A_2$ respectively, are anticommuting modulo the junk 2-forms of $A_3$.
\end{lemma}
\begin{demo}
One has $a_1[D,a_2]-[D,a_2]a_1=a_1[D,a_2]-[D,a_2a_1]+a_2[D,a_2]=0$, hence $[D,a_1][D,a_2]+[D,a_2][D,a_1]$ is a junk $2$-form for $A_3$. Moreover, from $[D,a_1a_2-a_2a_1]=0$ and $a_1[D,a_2]=[D,a_2]a_1$ one gets $a_2[D,a_1]=[D,a_1]a_2$. Then for any 1-forms $\omega_1=\sum a_1^i[D,b_1^i]$ and $\omega_2=\sum a_2^i[D,b_2^i]$, one has
\bea
\{\omega_1,\omega_2\}&=&\sum \left(a_1^i[D,b_1^i] a_2^j[D,b_2^j]+ a_2^j[D,b_2^j]a_1^i[D,b_1^i]\right)\cr
&=&\sum a_1^ia_2^j\{[D,b_1^i],[D,b_2^j]\}\in \mbox{junk of }A_3
\eea
\end{demo}
We can now express the curvature of $\hat\omega$ in terms of the curvatures of $\omega_{SM}$ and $\omega_\sigma$. We recall that all expressions involving curvature $2$-forms are modulo junk.

\begin{propo}\label{eq84}
We have $\rho(\hat \omega)=\rho(\omega_{SM})+\rho(\omega_{SM})^o+\rho(\omega_\sigma)$.
\end{propo}
\begin{demo}
We can apply  lemma \ref{lemsympa} to the case $A_3=\tilde{\cal C}^\infty_c(M,\hat \A_F^\ext)$, $A_1=\tilde{\cal C}^\infty_c(M,\hat\A_F)$ and $A_2=\tilde{\cal C}^\infty_c(M,\CC p_{\ell}\oplus \CC p_{q}\oplus\CC p_{\bar\ell}\oplus\CC p_{\bar q})$. Since  $\omega+\omega^o$ is a 1-form of $A_1$ and $\sigma$ is 1-form of $A_2$, we obtain that they anti-commute modulo junk. Thus
\bea
\rho(\hat\omega)&=&d\hat\omega+\hat\omega^2\cr
&=&\rho(\omega_{SM})+\rho(\omega_{SM})^o+\{\omega_{SM},\omega_{SM}^o\}+\rho(\omega_\sigma)+\{\omega_\sigma,\omega_{SM}+\omega_{SM}^o\}\cr
&=&\rho(\omega_{SM})+\rho(\omega_{SM})^o+\rho(\omega_\sigma)+\mbox{ junk, by lemmas \ref{curvomegaz} and }\ref{lemsympa}\nonumber
\eea
\ 
\end{demo}
Let us now compute $\rho(\omega_\sigma)$. To compute $d(i\gamma^\mu\hat\otimes Z_\mu' t_{B-L})$ we write $Z_\mu'=\sum a_i\partial_\mu b_i$, with $a_i,b_i$ some smooth functions. We obtain, using $[D,b_i\hat\otimes 1]=i\gamma^\mu\partial_\mu b_i\hat\otimes 1$:
\bea
d(i \gamma^\mu\hat\otimes Z_\mu't_{B-L})&=&d\left(\sum_i a_i\hat\otimes t_{B-L} [D,b_i\hat\otimes 1]\right)\cr
&=&\sum_i [D,a_i\hat\otimes t_{B-L}] [D,b_i\hat\otimes 1]\cr
&=&\sum_i[D_M,a_i][D_M,b_i]\hat\otimes t_{B-L}+\sum_ia_i\hat\otimes [D_F,t_{B-L}][D_M,b_i]\hat\otimes 1\cr
&=&-\sum_i \partial_\mu a_i\partial_\nu b_i\gamma^\mu\gamma^\nu\hat\otimes t_{B-L}-\sum_i a_i[D_M,b_i]\hat\otimes [D_F,t_{B-L}]\cr
&=&-{1\over 4}[\gamma^\mu,\gamma^\nu](\partial_\mu Z_\nu'-\partial_\nu Z_\mu')\hat\otimes t_{B-L}-i\gamma^\mu\hat\otimes [D_F, Z_\mu't_{B-L}]\nonumber
\eea
One computes that $[D_F,t_{B-L}]=\sigma(-2i)$. Hence we see that for any complex number $z'=x+iy$ we have $\sigma(z')={1\over 2}(xt_{B-L}[D_F,t_{B-L}]-y[D_F,t_{B-L}])$. We can use this decomposition to compute the differential of $1\hat\otimes\sigma(z')$:
\be
1\hat\otimes\sigma(z')={1\over 2}\left(x\hat\otimes t_{B-L})(1\hat\otimes[D_F,t_{B-L}]-(y\hat\otimes 1)(1\hat\otimes[D_F,t_{B-L}]\right)
\ee
Thus
\bea
d(1\hat\otimes \sigma(z'))&=&{1\over 2}\left([D_M,x]\hat\otimes t_{B-L} [D_F,t_{B-L}]+x\hat\otimes [D_F,t_{B-L}]^2-[D_M,y]\hat\otimes [D_F,t_{B-L}]\right)\cr
&=&i\gamma^\mu\hat\otimes \partial_\mu\sigma( z')+{1\over 2}x\hat\otimes [D_F,t_{B-L}]^2\label{finitediffsigma}
\eea
The second term is just the finite differential   $1\hat\otimes d_{D_F}\sigma(z')$ (see appendix \ref{projectionHiggs} for more details). We notice that if $m_0m_0^\dagger$ is a scalar matrix this term is in the junk. Now for the computation of $\omega_\sigma^2$. We have
\bea
\omega_\sigma^2&=&-\gamma^\mu\gamma^\nu Z_\mu' Z_\nu'\hat\otimes t_{B-L}^2+1\hat\otimes \sigma(z')^2+i\gamma^\mu \hat\otimes [Z_\mu't_{B-L},\sigma(z')] 
\eea
where we have used the rules for the graded tensor product. By the Clifford relations, the first term is a real function with values in $\hat\A_F^\ext$, hence it is in the junk. Gathering these results we obtain
\bea
\rho(\omega_\sigma)&=&-{1\over 4}[\gamma^\mu,\gamma^\nu]F_{\mu\nu}^{Z'}t_{B-L}+i\gamma^\mu\hat\otimes D_\mu \sigma(z)+1\hat\otimes  (\sigma(z')^2+d_F\sigma(z'))\cr
&:=&-{1\over 4}[\gamma^\mu,\gamma^\nu]F_{\mu\nu}^{Z'}t_{B-L}+i\gamma^\mu\hat\otimes D_\mu \sigma(z)+1\hat\otimes\rho_\sigma
\eea
where $F_{\mu\nu}^{Z'}=\partial_\mu Z_\nu'-\partial_\nu Z_\mu'$ is the curvature of the $Z'$-field, and the covariant derivative of the $\sigma$-field is defined by
\bea
D_\mu\sigma(z)&=&\partial_\mu\sigma(z')-[D_F+\sigma(z'),Z_\mu't_{B-L}]\cr
&=&\partial_\mu\sigma(z)+[Z_\mu't_{B-L},\sigma(z)]
\eea
A completely similar computation, which can be found in \cite{thesenadir}, yields the curvature of the SM part: 
%
\bea
\rho(\omega_{SM})&=&-{1\over 4}[\gamma^\mu,\gamma^\nu]\hat\otimes F_{\mu\nu}+i\gamma^\mu\hat\otimes D_\mu \Phi(q)+1\hat\otimes(\Phi(q')^2+d_F\Phi(q'))\cr
&:=&-{1\over 4}[\gamma^\mu,\gamma^\nu]\hat\otimes F_{\mu\nu}+i\gamma^\mu\hat\otimes D_\mu \Phi(q)+1\hat\otimes \rho_{Higgs}
\eea
where $F_{\mu\nu}=\partial_\mu A_\nu-\partial_\nu A_\mu+[A_\mu,A_\nu]$, and
\bea
D_\mu \Phi(q)&:=&\partial_\mu \Phi(q')-[D_F+\Phi(q'),A_\mu]\cr
&=&\partial_\mu \Phi(q)+[A_\mu,\Phi(q)]
\eea
We then obtain, using \eqref{formulesympa}:
\bea
\rho(\omega_{SM})^o&=&-{1\over 4}[\gamma^\mu,\gamma^\nu]^o\hat\otimes F_{\mu\nu}^o+(-1)(i\gamma^\mu)^o\hat\otimes (D_\mu \Phi(q))^o+1\hat\otimes((\Phi(q')^o)^2+d_F(\Phi(q')^o))\cr
&=&{1\over 4}[\gamma^\mu,\gamma^\nu]\hat\otimes F_{\mu\nu}^o+i\gamma^\mu\hat\otimes D_\mu \Phi(q)^o+1\hat\otimes\rho_{Higgs}^o\nonumber
\eea
where $D_\mu \Phi(q)^o:=\partial_\mu \Phi(q)^o-[A_\mu^o,\Phi(q)^o]$. Gathering all the terms we obtain:
\bea
\rho(\hat\omega)&=&-{1\over 4}[\gamma^\mu,\gamma^\nu]\hat\otimes\left(\FF_{\mu\nu}+F^{Z'}_{\mu\nu}t_{B-L}
\right)\cr
&&+i\gamma^\mu\hat\otimes D_\mu \Theta(q,z)+1\hat\otimes (\rho_{Higgs}+\rho_{Higgs}^o+\rho_\sigma)\label{curvature}
\eea
with $\Theta(q,z)=\Phi(q)+\Phi(q)^o+\sigma(z)$. We notice that all terms in \eqref{curvature} but the last are already orthogonal to the junk. The projection of $\rho_{Higgs}+\rho_{Higgs}^o+\rho_\sigma$ is computed in appendix \ref{projectionHiggs}. One finds
\bea
P(\rho_{Higgs})&=&-(|q|^2-1)[C_1,C_2,0,0]\cr
P(\rho_{Higgs}^o)&=&P(\rho_{Higgs})^o\cr
P(\rho_\sigma)&=&s(|z|^2-1)[D_1,0,D_1^*,0]
\eea
with
\bea
C_1&=&\colvec{\widetilde{Y_\nu^\dagger Y_\nu}&0\cr 0&\widetilde{Y_e^\dagger Y_e}}\oplus 1_3\otimes\colvec{ \widetilde{Y_u^\dagger Y_u}&0\cr 0&\widetilde{Y_d^\dagger Y_d}}\cr
C_2&=&\colvec{\widetilde{Y_\nu Y_\nu^\dagger}-{1\over k_\ell^2}\reel\tr(T_\ell Y_\nu Y_\nu^\dagger)T_\ell&0\cr 0&\widetilde{Y_e Y_e^\dagger}-{1\over k_\ell^2}\reel\tr(T_\ell Y_e Y_e^\dagger)T_\ell}\cr
&&\oplus 1_3\otimes\colvec{\widetilde{Y_u Y_u^\dagger}-{1\over k_q^2}\reel\tr(T_q Y_u Y_u^\dagger)T_q&0\cr 0&\widetilde{Y_d Y_d^\dagger}-{1\over k_q^2}\reel\tr(T_q Y_d Y_d^\dagger)T_q}\cr
D_1&=&\colvec{\widetilde{m_0^\dagger m_0}&0\cr 0&0}
\eea
\subsection{The bosonic action}
There now only remains to calculate the Krein-Schmidt squared  norm of $P(\rho(\hat\omega))$. From the previous section we have $P(\rho(\hat\omega))=R_0+R_1+R_2$, with
\bea
R_0&=& 1\hat\otimes (P(\rho_{Higgs})+P(\rho_{Higgs})^o+P(\rho_\sigma))\cr
R_1&=&i\gamma^\mu\hat\otimes D_\mu \Theta(q,z)\cr
R_2&=&-{1\over 4}[\gamma^\mu,\gamma^\nu]\hat\otimes\left(\FF_{\mu\nu}+F^{Z'}_{\mu\nu}t_{B-L}\right)
\eea
The 3 terms $R_{0,1,2}$ are orthogonal to each other thanks to the properties of the Hodge product on forms. Let us focus first on $R_0$. The trace over $\K_F$ of the square of the projected finite curvature is computed in appendix \ref{projectionHiggs}. Here we must also trace the identity matrix over the Dirac spinor space, yielding an additional factor of four. We thus obtain (compare with equation \eqref{normesq}):
\bea
(R_0,R_0)_\RR&=&8V_0(|q|^2-1)^2+8W_0(|z|^2-1)^2-16sK(|q|^2-1)(|z|^2-1)\cr
&&
\eea
where
\bea
V_0&=&\tr(C_1^2+ C_2^2)\cr
&=&\|\widetilde{Y_\nu Y_\nu^\dagger}\|^2+\|\widetilde{Y_e Y_e^\dagger}\|^2+3\|\widetilde{Y_u Y_u^\dagger}\|^2+3\|\widetilde{Y_d Y_d^\dagger}\|^2\cr
&&+2{\|\widetilde{Y_\nu Y_\nu^\dagger}\|^2\|\widetilde{Y_e Y_e^\dagger}\|^2\over \|\widetilde{Y_\nu Y_\nu^\dagger}-\widetilde{Y_e Y_e^\dagger}\|^2}\sin^2(\theta_\ell)+6{\|\widetilde{Y_u Y_u^\dagger}\|^2\|\widetilde{Y_d Y_d^\dagger}\|^2\over \|\widetilde{Y_u Y_u^\dagger}-\widetilde{Y_d Y_d^\dagger}\|^2}\sin^2(\theta_q)\cr
W_0&=&\tr(D_1^2)=\|\widetilde {m_0m_0^\dagger}\|^2\cr
K&=&\reel \tr(C_1D_1)=\reel\tr(\widetilde{Y_\nu^\dagger Y_\nu}\widetilde{m_0^\dagger m_0})
\eea
where the angles $\theta_\ell$ and $\theta_q$ are defined up to sign by
\bea
(\widetilde{Y_\nu Y_\nu^\dagger},\widetilde{Y_e Y_e^\dagger})_\RR=\|\widetilde{Y_\nu Y_\nu^\dagger}\|\|\widetilde{Y_e Y_e^\dagger}\|\cos(\theta_\ell)\cr
(\widetilde{Y_u Y_u^\dagger},\widetilde{Y_d Y_d^\dagger})_\RR=\|\widetilde{Y_u Y_u^\dagger}\|\|\widetilde{Y_d Y_d^\dagger}\|\cos(\theta_q).
\eea
\begin{rem}
In the (non-extended) NCSM, $V_0$ has the same value as above except that the two last terms containing sines are suppressed.
\end{rem}

Let us now look at $R_2$. We have 
\bea
(R_2,R_2)&=&{1\over 16}\tr([\gamma^\mu,\gamma^\nu]^\times[\gamma^\lambda,\gamma^\rho])\tr((\FF_{\mu\nu}+F^{Z'}_{\mu\nu}t_{B-L})^\times (\FF_{\lambda\rho}+F^{Z'}_{\lambda\rho}t_{B-L}))\cr
&=&(-g^{\mu\rho} g^{\nu\lambda}+g^{\mu\lambda}g^{\nu\rho})\tr((\FF_{\mu\nu}+F^{Z'}_{\mu\nu}t_{B-L})^\times (\FF_{\lambda\rho}+F^{Z'}_{\lambda\rho}t_{B-L}))\cr
&=&2\tr((\FF_{\mu\nu}+F^{Z'}_{\mu\nu}t_{B-L})^\times  (\FF^{\mu\nu}+F^{Z'\mu\nu}t_{B-L}))
\eea
To compute these terms, we write:
\be
\FF_{\mu\nu}=\FF_{\mu\nu}^Yt_Y+\FF^W_{\mu\nu a}t_W^a+\FF^C_{\mu\nu a}t_C^a.
\ee
Thanks to the scalar products obtained in section \ref{secextab} we find that 
\bea
(\FF_{\mu\nu}+F_{\mu\nu}^{Z'}t_{B-L},\FF_{\lambda\rho}+F^{Z'}_{\lambda\rho}t_{B-L})&=&{80\over 3}N\FF_{\mu\nu}^Y\FF_{\lambda\rho}^Y+16N\FF_{\mu\nu a}^W\FF_{\lambda\rho a}^W\cr
&&+16N\FF_{\mu\nu a}^C\FF_{\lambda\rho a}^C+{32\over 3}NF_{\mu\nu}^{Z'}F_{\lambda\rho}^{Z'}\cr
&&+{64\over 3}N\FF_{\mu\nu}^Y F_{\lambda\rho}^{Z'}\label{kineticterm}
\eea
The last term, known as kinetic mixing, is a generic feature of $U(1)'$ extensions of the SM \cite{holdom}. Thus we obtain:
\bea
(R_2,R_2)&=&2({80\over 3}N\FF_{\mu\nu}^Y\FF^{Y\mu\nu}+16N\FF_{\mu\nu a}^W\FF^{W\mu\nu a}+16N\FF_{\mu\nu a}^C\FF^{C\mu\nu a}\cr
&&+{32\over 3}NF_{\mu\nu}^{Z'}F^{Z'\mu\nu}+{64\over 3}N\FF_{\mu\nu}^Y F^{Z'\mu\nu})
\eea

Let us now look at $R_1$:
\bea
(R_1,R_1)&=&(i\gamma^\mu\hat\otimes D_\mu \Theta,i\gamma^\nu\hat\otimes D_\nu \Theta)\cr
&=&(\gamma^\mu\hat\otimes D_\mu \Theta,\gamma^\nu\hat\otimes D_\nu \Theta)\cr
&=&\tr((\gamma^\mu\hat\otimes D_\mu \Theta)^\times \gamma^\nu\hat\otimes D_\nu \Theta)\cr
&=&-\tr(((\gamma^\mu)^\times\hat\otimes D_\mu \Theta^\times) \gamma^\nu\hat\otimes D_\nu \Theta),\mbox{ since }(A\hat\otimes B)^\times=(-1)^{|A||B|}A^\times\hat\otimes B^\times \cr
&=&-\tr((\gamma^\mu\hat\otimes D_\mu \Theta) \gamma^\nu\hat\otimes D_\nu \Theta)\cr
&=&\tr(\gamma^\mu\gamma^\nu\hat\otimes D_\mu \Theta D_\nu \Theta)\cr
&=&\tr(\gamma^\mu\gamma^\nu)\tr(D_\mu \Theta D_\nu \Theta)\cr
&=&4g^{\mu\nu}\tr(D_\mu \Theta D_\nu \Theta)\cr
&=&4\tr(D_\mu \Theta D^\mu \Theta)
\eea
We now observe that $D_\mu\Phi D_\nu\Phi^o=0$. Moreover $D_\mu \Phi D_\nu \sigma$ is off-diagonal, hence traceless. Thus we obtain
\bea
(R_1,R_1)&=&4\left(2\tr(D_\mu\Phi(q)D^\mu\Phi(q))+\tr(D_\mu\sigma(z)D^\mu\sigma(z)\right))
\eea
The generalized Connes-Lott   bosonic Lagrangian for the extended SM is thus:
\bea
{\cal L}_b&=&-{160\over 3}N\FF_{\mu\nu}^Y\FF^{Y\mu\nu}-{32}N\FF_{\mu\nu a}^W\FF^{W\mu\nu a}-{32}N\FF_{\mu\nu a}^C\FF^{C\mu\nu a}\cr
&&-{64\over 3}NF_{\mu\nu}^{Z'}F^{Z'\mu\nu}-{128\over 3}N\FF_{\mu\nu}^Y F^{Z'\mu\nu}\cr
&&-8\tr(D_\mu\Phi(q)D^\mu\Phi(q))-4\tr(D_\mu\sigma(z)D^\mu\sigma(z))-V(q,z)\label{bosoniclag}
\eea
%
%
where 
\be
V(q,z)=8V_0(|q|^2-1)^2+8W_0(|z|^2-1)^2-16sK(|q|^2-1)(|z|^2-1)
\ee
Note that $V(q,z)$ is non-negative by definition (see appendix \ref{projectionHiggs}). The sign in front of the kinetic term of $\Phi(q)$ and $\sigma(z)$ may look suspicious, but we will see in the next section that it yields the correct sign for the kinetic term of the Higgs and complex scalar when we develop the matrix products, the fundamental reason being that $\Phi(q)$ and $\sigma(z)$ are Krein selfadjoint and not Hilbert selfadjoint. Hence at this point the non-triviality of the finite metric $\eta_F$ plays a fundamental role. It is striking that it yields all the correct signs for both the bosonic and the fermionic Lagrangians.

We see that we obtain all the gauge invariant terms of the SM coupled with a B-L Z' boson and a complex scalar. What is gained is that the relative signs of the couplings are fixed, as well as the form of the quartic potential of the Higgs, complex scalar fields and the particular form of the coupling between them (where all fourth degree polynomials in $|z|$ and $|q|$ would be gauge-invariant).

\section{Identification of the physical fields}
  

We now need to identify the physical fields appearing in \eqref{bosoniclag}. For this we develop the kinetic terms. We use the same notations as \cite{VS}.

We   obtain:
\be
\tr(D_\mu\Phi(q)D^\mu\Phi(q))=-2a|D_\mu   H|^2 \label{trdmu}
\ee
where 
\be 
a=\tr(Y_e Y_e^\dagger+Y_\nu Y_\nu^\dagger+3M_u+3M_d),
\ee
$H$ is the first column of the quaternion $q$,  $D_\mu$ is the operator
\be
D_\mu=\partial_\mu +i\BB_{\mu,a}^W\sigma^a-i\BB_\mu^Y,\label{covder}
\ee
and $|D_\mu H|^2=|D_0H|^2-\sum_{i=1}^3|D_i H|^2$.

\begin{rem}
We see from \eqref{covder} that $H$ has hypercharge $-1$ and weak isospin $1/2$. It is the conjugate of the Higgs field. Clearly $|D_\mu H|=|D_\mu^* H^*|$ with $D_\mu^*=\partial_\mu -i\BB_{\mu,a}^W(\sigma^a)^*+i\BB_\mu^Y$, so we could have written \eqref{trdmu} in terms of the Higgs field $H^*$, but it is more natural to use $H$ which is a column of $q$.
\end{rem} 

We must now compute the kinetic term for the $\sigma$-field. We find:
\bea
D_\mu\sigma(z)&=&\partial_\mu \sigma(z)+[Z_\mu't_{B-L},\sigma(z)]\cr
&=&\sigma(\partial_\mu z+2iZ_\mu'z)\cr
&:=&\sigma(D_\mu z)
\eea
From which we obtain 
\bea
\tr(D_\mu \sigma(z)D^\mu\sigma(z))&=&2bs|D_\mu z|^2 
\eea
with $b=\tr(m_0m_0^\dagger)$ and $D_\mu z=\partial_\mu z+2iZ_\mu'z$. Introducing this into   \eqref{bosoniclag}, we get:
\bea
{\cal L}_b &=&-160{N\over 3}\FF_{\mu\nu}^Y\FF^{Y\mu\nu}-{32}N\FF_{\mu\nu a}^W\FF^{W\mu\nu a}-{32}N\FF_{\mu\nu a}^C\FF^{C\mu\nu a}\cr
&&-{64\over 3}NF_{\mu\nu}^{Z'}F^{Z'\mu\nu}-{128\over 3}N\FF_{\mu\nu}^Y F^{Z'\mu\nu}\cr
&&+16a|D_\mu H|^2-8bs|D_\mu z|^2\cr
&&-8V_0(|H|^2-1)^2-8W_0(|z|^2-1)^2+16sK(|H|^2-1)(|z|^2-1)
\eea
We see that in order to have the correct sign in front of the kinetic term for $z$ we must suppose $s=-1$. We will do it from now on. Let us introduce the normalized fields $Y,W,G,\tilde H$ and $\hat Z'$:
\bea
\BB_\mu^Y={1\over 2}g_YY_\mu,&&\BB_\mu^{Wa}={1\over 2}g_w W_\mu^a \cr
\BB_\mu^{Ca}={1\over 2}g_sG_\mu^a, &&  Z_\mu'={1\over 2}g_{Z'}\hat Z_\mu'\cr
 H=k \tilde H,&&   z= l \tilde z
\eea
The constants are defined in order to obtain normalized kinetic terms\footnote{The normalizations are the same as in Peskin-Schroder /Langacker. To obtain Weinberg normalization just replace $\tilde H$ with $\sqrt{2}\tilde H$}:
\bea
{\cal L}_b^{SM}&=&-{1\over 4}|Y_{\mu\nu}|^2 -{1\over 4}|W_{\mu\nu}^a|^2-{1\over 4}|G_{\mu\nu}^a|^2-{1\over 4}|\hat Z_{\mu\nu}'|^2-{\kappa \over 2}Y_{\mu\nu}\hat {Z'}^{\mu\nu}\cr 
&&+ |D_\mu \tilde H|^2+ |D_\mu \tilde z|^2-V(\tilde H,\tilde z)
\eea
Hence
\bea
g_w^2=g_s^2={5\over 3}g_Y^2={2\over 3}g_{Z'}^2={1\over 32N} ,&& \kappa=64 {N\over 3}g_Yg_{Z'}=\sqrt{2\over 5}\cr
k^2={1\over 16a}, &&l^2={1\over 8b} \label{couplings1}
\eea
To deal with the kinetic mixing term we must change basis in the $(Y,Z')$-space and several choices are possible. We can do a $\pi/4$-rotation followed by a normalization \cite{CRM} or a triangular transformation \cite{salvioni} to obtain new fields $\tilde Y$ and $\tilde Z$. In the first case we do
\be 
\colvec{Y\cr \hat Z'}={\sqrt{2}\over 2}\colvec{{1\over \sqrt{1+\kappa}}&-{1\over \sqrt{1-\kappa}}\cr {1\over \sqrt{1+\kappa}}&{1\over \sqrt{1-\kappa}}}\colvec{\tilde Y\cr \tilde Z'}
\ee
and in the second case
\be 
\colvec{Y\cr \hat Z'}= \colvec{1&-{\kappa\over \sqrt{1-\kappa^2}}\cr 0&{1\over \sqrt{1-\kappa^2}}}\colvec{\tilde Y\cr \tilde Z'}\label{triang}
\ee
In both cases the change of basis depends on $\kappa$, hence on the coupling constants, and will not be invariant under the renormalization flow. To understand the coupling constant of the new field, we can look at the part of the Dirac operator which depends on them, since it is the Dirac operator which gives the interactions between fermions. We have (dropping the overall factor ${1\over 2}\gamma^\mu\hat\otimes$):
\bea
g_YY_\mu t_Y+g_{Z'}\hat Z_\mu't_{B-L}&=&g_Y\tilde Y_\mu t_Y+\tilde Z_\mu'({-\kappa\over \sqrt{1-\kappa^2}}g_Yt_Y+{1\over\sqrt{1-\kappa^2}}g_{Z'}t_{B-L})\cr
&:=&g_Y\tilde Y_\mu t_Y+\tilde Z_\mu'(\tilde gt_Y+g't_{B-L})\label{defgprimegtilde}
\eea
where we have used the same notations as in \cite{CRM}.

\begin{rem}
The procedures just described are natural in a setting where one just plugs in all the gauge-invariant terms in the Lagrangian. However in the NCG setting, it would be natural to change basis right from the start. Removing the orthogonal projection of $t_{B-L}$ onto $t_Y$, one obtains the vector
\bea
t_{Z'}&:=&t_{B-L}-(t_{B-L},t_Y){t_Y\over \|t_Y\|^2}\cr
&=&t_{B-L}-{2\over 5}t_Y\cr
&=&[\colvec{-i&0\cr 0&-{1\over 5}i}\oplus \colvec{-{1\over 5}i&0\cr 0&{3\over 5}i}\otimes 1_3,{-3\over 5}i1_2\oplus {1\over 5}i1_2\otimes 1_3,c,c]\otimes 1_N
\eea
which is orthogonal to $t_Y$. Its squared norm is $-\tr(t_{Z'}^2)={32N\over 5}$. Hence we define the new field components\footnote{The basis is of course more important than the components since it is the basis which allows to physically interpret the fields. Hence, even if $\bat Z'$ and $Z'$ have the same components, they are different fields, with different charges.} $\BB_\mu^{\bat Y}=\BB_\mu^Y+{2\over 5}Z_\mu'$ and $\bat Z_\mu'=Z_\mu'$ such that $\BB_\mu^{\bat Y}t_Y+\bat Z_\mu't_{Z'}=\BB_\mu^Yt_Y+Z_\mu't_{B-L}$. Since the curvature is linear in the abelian fields, one also gets   $\FF_{\mu\nu}^{\bat Y}t_Y+F_{\mu\nu}^{\bat Z'}t_{Z'}=\FF_{\mu\nu}^Yt_Y+F_{\mu\nu}^{Z'}t_{B-L}$. Redoing the computation of \eqref{kineticterm}, we now find
\bea
{\cal L}_b^{gauge}&=&-{160\over 3}N\FF_{\mu\nu}^{\bat Y}\FF^{\bat Y\mu\nu}-{32}N\FF_{\mu\nu a}^W\FF^{W\mu\nu a}-{32}N\FF_{\mu\nu a}^C\FF^{C\mu\nu a}-{64N\over 5}F_{\mu\nu}^{\bat Z'}F^{\bat Z'\mu\nu}\cr
\eea
writing $\BB^{\bat Y}={1\over 2}g_Y\tilde Y$ and $\bat Z'={1\over 2}g_{\bat Z'}\tilde Z'$, the kinetic term for the gauge fields is normalized with the same values of the $g_Y,g_w,g_s$, and $g_{\bat Z'}^2={5\over 64 N}$, so that the following relation holds:
\be 
 g_w^2=g_s^2={5\over 3}g_Y^2={2\over 5}g_{\bat Z'}^2 
\ee
This way of removing the kinetic mixing is easily shown to be completely equivalent to \eqref{triang}.
\end{rem}

Let us now look at the scalar sector. The potential is (using the same notations as \cite{BMP}):
\bea
V(\tilde H,\tilde z)&=&8V_0(k^2|\tilde H|^2-1)^2+8W_0(l^2|\tilde z|^2-1)^2+16K(k^2|\tilde H|^2-1)(l^2|\tilde z|^2-1)\cr
&=&m_1^2|\tilde H|^2+m_2^2|\tilde z|^2+\lambda_1 |\tilde H|^4+\lambda_2|\tilde z|^4+\lambda_3|\tilde H|^2|\tilde z|^2+\mu 
\eea
where
\bea
\lambda_1&=&8k^4V_0={V_0\over 32 a^2}\cr
\lambda_2&=&8l^4W_0={W_0\over 8b^2}\cr
\lambda_3&=&16Kk^2l^2={K\over 8 ab}\cr
m_1^2&=&-16k^2(V_0+K)=-{V_0+K\over a}\cr
m_2^2&=&-16l^2(W_0+K)=-2{W_0+K\over b}\cr
\mu&=&8V_0+8W_0+16K\label{couplings2}
\eea
By construction, the minimal of the potential is zero, since it is originally of the form $\tr(A^2)$, with $A$ some matrix (this is one of the advantages of the Connes-Lott approach). It is thus obtained for $|H|=|z|=1$, which correspond to 
\be 
|\tilde H|^2={1\over k^2}={ 16 a}:={v^2\over 2},\quad |\tilde z|^2={1\over l^2}={ 8b}:={(v')^2\over 2}
\ee
Using gauge invariance we bring $\tilde H$ and $\tilde z$ into the form $\tilde H=\colvec{|\tilde H|\cr 0}$ and $\tilde z=|\tilde z|$ and expand around the minimum, defining the real fields $h$ and $h'$ such that $|\tilde H|={1\over\sqrt{2}}(v+h)$ and $|\tilde z|={1\over\sqrt{2}}(v'+h')$. The quadratic term\footnote{Note that in the Connes-Lott approach there is no constant term, hence no contribution to the cosmological constant.} comes out as
\be 
q(h,h')=\lambda_1v^2 h^2+\lambda_2 {v'}^2 {h'}^2+\lambda_3vv'hh'\label{quadra}
\ee
%


To compute the masses of the scalar fields we must move to a basis where this quadratic form is diagonal.  We write
\be
\colvec{h_1\cr h_2}=\colvec{\cos \alpha&-\sin \alpha\cr \sin \alpha&\cos\alpha}\colvec{h\cr h'}
\ee
such that 
\be 
q(h_1,h_2)={1\over 2}m_{h_1}^2h_1^2+{1\over 2}m_{h_2}^2h_2^2
\ee
computing the eigenvalues   of the matrix of the quadratic form \eqref{quadra}, we find (after many others, see \cite{BMP}, \cite{CRM}):
\bea
m_{h_1/h_2}^2&=&\lambda_1v^2+\lambda_2{v'}^2\mp\sqrt{(\lambda_1v^2-\lambda_2{v'}^2)^2+(\lambda_3vv')^2}\cr
&=&{V_0\over a}+{2W_0\over b}\mp \sqrt{({V_0\over a}-{2W_0\over b})^2+{8K^2\over ab}} 
\eea
From $q(h,h')=q(h_1,h_2)$ we then see that $2\alpha$ satisfies
\bea
\cos(2\alpha)&=&{\lambda_2{v'}^2-\lambda_1v^2\over \sqrt{(\lambda_1v^2-\lambda_2{v'}^2)^2+(\lambda_3vv')^2}}\cr
\sin(2\alpha)&=&{\lambda_3vv'\over \sqrt{(\lambda_1v^2-\lambda_2{v'}^2)^2+(\lambda_3vv')^2}}
\eea
We can invert these relations to obtain the parameters of the Lagrangian in terms of mass and mixing angle\footnote{Note a sign problem  in \cite{CRM} eq. 41, which is inconsistent with eq 40. There is the same problem in \cite{BMP}, eq 13,14}:
\bea
\lambda_1&=&{1\over 4v^2}(m_{h_1}^2(1+\cos(2\alpha))+m_{h_2}^2(1-\cos(2\alpha))\cr
\lambda_2&=&{1\over 4{v'}^2}(m_{h_1}^2(1-\cos(2\alpha))+m_{h_2}^2(1+\cos(2\alpha))\cr
\lambda_3&=&{m_{h_2}^2-m_{h_1}^2\over 2vv'}\sin(2\alpha)
\eea
Let us now look at kinetic term of the scalar fields. We have
\bea
D_\mu\tilde H&=&(\partial_\mu +i\BB_{\mu,a}^W\sigma^a-i\BB_\mu^Y)\tilde H\cr
&=&(\partial_\mu +{1\over 2}ig_w W_{\mu,a}\sigma^a-{1\over 2}ig_Y Y_\mu)\tilde H\cr
&=&(\partial_\mu +{1\over 2}ig_w W_{\mu,a}\sigma^a-{1\over 2}ig_Y (\tilde Y_\mu-{\kappa\over \sqrt{1-\kappa^2}})\tilde Z_\mu')\tilde H\cr
&=&(\partial_\mu +{1\over 2}ig_w W_{\mu,a}\sigma^a-{1\over 2}ig_Y \tilde Y_\mu-{1\over 2}i\tilde g\tilde Z_\mu')\tilde H
\eea

Similarly,
\bea
D_\mu \tilde z&=&\partial_\mu \tilde z+ig_{Z'}\hat Z_\mu' \tilde z\cr
&=&\partial_\mu \tilde z+ig'\tilde Z_\mu' \tilde z
\eea
The kinetic term is thus
\bea
|D_\mu \tilde H|^2+|D_\mu \tilde z|^2&=&{1\over 2}|\partial_\mu h|^2+{1\over 2}|\partial_\mu h'|^2+{1\over 8}(v+h)^2g_w^2|W_{\mu,1}+iW_{\mu,2}|^2\cr
&&+{1\over 8}(v+h)^2|g_wW_{\mu,3}-g_Y\tilde Y_\mu-\tilde g\tilde Z_\mu'|^2+{1\over 2}{g'}^2(v'+h')^2|\tilde Z_\mu'|^2\cr
\eea
The $W$-bosons  are not affected by the extension. Charge eigenstates $W_\mu^\pm$ are introduced as usual and their tree-level mass is 
\be 
M_W={1\over 2}vg_w\label{massW}
\ee

For the remaining fields, the mass term to be diagonalized is thus:
\bea
q(\tilde Y_\mu,W_{\mu,3},\tilde Z_\mu)&=&{1\over 8}(v^2|g_wW_{\mu,3}-g_Y\tilde Y_\mu-\tilde g\tilde Z_\mu'|^2+4{v'}^2{g'}^2|\tilde Z_\mu'|^2)
\eea
We first rotate the orthogonal basis $(t_Y,t_Z,t_{Z'})$ around $t_Z$ of an angle $\theta_w$ (weak mixing angle) in order to identify the photon and $Z$ states. The transformation is thus $\colvec{A_\mu\cr Z_\mu\cr \tilde Z_\mu'}=\colvec{\cos\theta_w&\sin\theta_w&0\cr -\sin\theta_w&\cos\theta_w&0\cr 0&0&1}\colvec{\tilde Y_\mu\cr W_{\mu,3}\cr \tilde Z_\mu'}$ with $\tan\theta_w={g_Y\over g_w}$. The quadratic form becomes
\bea
q(A_\mu,Z_\mu,\tilde Z_\mu')&=&{1\over 8}(v^2|\sqrt{g_w^2+g_Y^2}Z_\mu-\tilde g\tilde Z_\mu'|^2+4{g'}^2{v'}^2|\tilde Z_\mu'|^2)
\eea
Now we do a second rotation around the photon axis (since the photon must remain massless). Defining the $ZZ'$-mixing angle\footnote{This angle is already constrained by experimental data to be less than $0.01$, according to \cite{andreev}, and ignoring kinetic mixing.} $\theta'$ and the new fields $Z_\mu^{\rm new},Z_\mu^{'\rm new}$, the transformation is thus
\bea
\colvec{A_\mu\cr Z_\mu^{\rm new}\cr  Z_\mu^{'\rm new}}&=&\colvec{1&0&0\cr 0&\cos\theta'&\sin\theta'\cr 0&-\sin\theta'&\cos\theta'}\colvec{A_\mu\cr Z_\mu \cr \tilde Z_\mu'}
\eea
and $\theta'$ satisfies\footnote{At this point it is useful to know that the angle of rotation for the diagonalization of the symmetric matrix $\colvec{a&c\cr c&b}$ satisfies $\tan\alpha={2c\over a-b}$.}:
\be 
\tan(2\theta')={2\tilde g\sqrt{g_w^2+g_Y^2}\over \tilde g^2+4{g'}^2\left({v'\over v}\right)^2-g_w^2-g_Y^2}
\ee
The masses of $Z$ and $Z'$ squared are the eigenvalues of 
$${1\over 4}\colvec{v^2(g_w^2+g_Y^2)&-v^2\tilde g\sqrt{g_w^2+g_Y^2}\cr -v^2\tilde g\sqrt{g_w^2+g_Y^2}&\tilde g^2v^2+4{g'}^2{v'}^2}$$
 
The masses of the different gauge bosons thus satisfy (at unification scale):
\bea
M_W^2&=&{1\over 4}v^2g_w^2\cr
M_Z^2+M_{Z'}^2&=&{1\over 4}(g_w^2v^2+g_Y^2v^2+\tilde g^2v^2+4{g'}^2{v'}^2)\cr
M_Z^2M_{Z'}^2&=&{1\over 4}v^2{v'}^2(g_w^2+g_Y^2){g'}^2\label{massbosons1}
\eea
One can then solve for $v$ and $v'$ in the first and last equations, and use the solution in the expression for the Higgs masses. The second equation is then seen as a relation between gauge bosons masses. We get
\bea
M_Z^2+M_{Z'}^2&=&{g_w^2+g_Y^2+\tilde g^2\over g_w^2}M_W^2+{g_w^2\over g_w^2+g_Y^2}{M_Z^2M_{Z'}^2\over M_W^2}\cr
m_{h_1}^2+m_{h_2}^2&=&{4M_W^2\over g_w^2}\lambda_1 +{M_Z^2M_{Z'}^2\over M_W^2}{g_w^2\over (g_w^2+g_Y^2){g'}^2}\lambda_2\cr
m_{h_1}^2m_{h_2}^2&=&(\lambda_1\lambda_2-{1\over 4}\lambda_3){4M_Z^2M_{Z'}^2\over (g_w^2+g_Y^2){g'}^2}\label{massbosons2}
\eea

\begin{rem}
The mass eigenvalues for the $Z$ and $Z'$ bosons have quite complicated expressions. In order to write them down, let us introduce \cite{salvioni}:
\bea
g_Z&=&\sqrt{g_w^2+g_Y^2}\cr
M_{Z'}^0&=&{1\over 2}\sqrt{ \tilde g^2v^2+4{g'}^2{v'}^2}\cr
M_Z^0&=&{1\over 2}vg_Z
\eea
where $M_Z^0$ and $M_{Z'}^0$ would be the masses of the $Z$ and $Z'$ bosons in the absence of kinetic mixing. Then one has:
\be 
M_{Z/Z'}^2={1\over 2}(M_Z^0)^2\left[1+\left({M_{Z'}^0\over M_Z^0}\right)^2\mp\sqrt{\left(1-\left({M_{Z'}^0\over M_Z^0}\right)^2\right)^2+4{\tilde g^2\over g_Z^2}}\right]
\ee
which in terms of coupling constants and vev gives
\bea
M_{Z/Z'}^2&=&{1\over 8}\left(g_w^2v^2+g_Y^2v^2+\tilde g^2v^2+4{g'}^2{v'}^2\mp \sqrt{(g_w^2v^2+g_Y^2v^2-\tilde g^2v^2-4{g'}^2{v'}^2)^2+4\tilde g^2(g_w^2+g_Y^2)v^4} \right)\cr
&=&{1\over 8}\left(g_w^2v^2+g_Y^2v^2+\tilde g^2v^2+4{g'}^2{v'}^2\mp \sqrt{(g_w^2v^2+g_Y^2v^2+\tilde g^2v^2-4{g'}^2{v'}^2)^2+16\tilde g^2{g'}^2v^2{v'}^2} \right)\nonumber
\eea
\end{rem}
 

\section{The fermionic action}
Traditionally, the fermionic action is taken to be\footnote{We use the Lorentzian definition here. In the Euclidean case, the action is $S_f(\omega,\psi)={1\over 2}\bra J\psi,D_\omega\psi\ket$.} 
\be 
S_f(\omega,\psi)={1\over 2}(\psi,D_\omega\psi)\label{usualfa}
\ee
where $D_\omega=D+\omega+\omega^o$ and $\psi$ is a Grassmann field with values in $S\hat\otimes \K_F$. Since $S\hat\otimes \K_F$ has four times too many degrees of freedom, there is a fermion quadrupling problem which can be taken care of by imposing the Majorana-Weyl conditions on $\psi$:
\bea
\chi\psi&=&\psi\cr
J\psi&=&\psi\label{MW}
\eea
It has been shown in \cite{barrettunique} that, apart from a phase which we ignore here, \eqref{MW} defines the only real subspace of the correct physical dimension which is invariant under the symmetry group of the fermionic action. However, it has also been observed that instead of restricting to this subspace one can start with the symmetrical fermionic action
\begin{eqnarray}
S_{sym}(\omega,\psi)=&\frac{1}{16}[(\psi,D_\omega\psi)+(D_\omega\psi,\psi)+(D_\omega\psi,J\psi)+(J\psi,D_\omega\psi)\cr
&+(\chi\psi,D_\omega\psi)+(D_\omega\psi,\chi\psi)+(D_\omega\psi,\chi J\psi)+(\chi J\psi,D_\omega\psi)]
\end{eqnarray}
which can be rewritten
\be 
S_{sym}(\omega,\psi)=\frac{1}{2}(\pi\psi,D_\omega \pi\psi)\label{symaction}
\ee
where $\pi={1\over 4}(1+J)(1+\chi)$, and using $\epsilon=\epsilon''=1$, $\kappa=\kappa''=-1$ and $(\phi,A\psi)=-(\psi,A^\times \phi)$ for an anti-linear operator $A$. Now $\pi$ is the Krein selfadjoint projector on the space defined by \eqref{MW}, so that using the action \eqref{symaction} is in effect equivalent to using \eqref{usualfa} with $\psi$ submitted to \eqref{MW}. Seeing $\psi$ as a field with values in $S\otimes \K_0\otimes \CC^4$, one sees that
\be
\psi=\sum_p\psi_R^p\otimes p\otimes R+J_M\psi_R^p\otimes p\otimes \bar R+\psi_L^p\otimes p\otimes L-J_M\psi_L^p\otimes p\otimes \bar L
\ee
where $p$ runs through the elementary particle orthonormal basis $\nu_i,e_i,u_i^j,d_i^j$ where $i$ and $j$ are  the generation and color indices, respectively. 

We recall that 
\bea
D_\omega&=&(i\gamma^\mu\partial_\mu\hat\otimes 1)+(i\gamma^\mu\hat\otimes (\BB_\mu+Z_\mu't_{B-L}))+(1\hat\otimes(\Phi(q)+\Phi(q)^o+\sigma(z))\nonumber
\eea
where the bracketing yields the decomposition of the fermionic action into a kinetic, gauge and Higgs part, which we will compute separately.
 
The ``ket-bra'' notation 
\bea
\Phi(q)&=&Y(q)\otimes |L\ket\bra R|-Y(q)^\dagger\otimes |R\ket\bra L|\cr
\Phi(q)^o&=&-Y(q)^T\otimes |\bar R\ket\bra \bar L|+\bar Y(q)\otimes |\bar L\ket\bra\bar R|\cr
\sigma(z)=\sigma^o(z)&=&zM_0\otimes |\bar R\ket\bra R|-z^*M_0^\dagger\otimes |R\ket\bra\bar R|
\eea
where $Y(q)=\tilde q Y_0$, will be useful.

For the kinetic part, we obtain the usual expression
\bea
S_f^{kin}&=&\sum_p\big((\psi_R^p,i\gamma^\mu\partial_\mu\psi_R^p)+(\psi_L^p,i\gamma^\mu\partial_\mu \psi_L^p)\big)
\eea
Note that it is crucial in this calculation that the internal metric be $[1,-1,-1,1]$.

We now compute the Higgs part of the fermionic action. We note that
\bea
(\psi,1\hat\otimes \Phi(q)^o\psi)&=&( \psi,1\hat\otimes J_F\Phi(q)^\times J_F^{-1} \psi)\cr
&=&( \psi,J(1\hat\otimes  \Phi(q)^\times) J^{-1} \psi)\cr
&=&(1\hat\otimes \Phi(q)^\times J^{-1}\psi,J\psi)\cr
&=&((1\hat\otimes \Phi(q))^\times  \psi,\psi)\cr
&=&(\psi,1\hat\otimes \Phi(q)\psi)
\eea
Now, we have, using $1\hat\otimes \Phi(q)=\chi_M\otimes \Phi(q)$:
\bea
(\psi,1\hat\otimes \Phi(q) \psi)&=&(\psi,\chi_M\otimes (Y(q)\otimes |L\ket\bra R|-Y(q)^\dagger\otimes |R\ket\bra L|)\psi)\cr
&=&\sum_{p,p'}(\psi_L^p\otimes p\otimes L,\psi_R^{p'}\otimes Y(q)p'\otimes L)+\sum_{p,p'}(\psi_R^p\otimes p\otimes R,\psi_L^{p'}\otimes Y(q)^\dagger p'\otimes R)\cr
&=&\sum_{p,p'}\big((\psi_L^p,\psi_R^{p'})\bra p,Y(q)p'\ket+(\psi_R^p,\psi_L^{p'})\bra p,Y(q)^\dagger p'\ket\big)\cr
&=&\sum_{i,i'}\big(\alpha(\psi_L^{\nu_i},\psi_R^{\nu_{i'}})(Y_\nu)_{ii'}+\beta(\psi_L^{\nu_i},\psi_R^{e_{i'}})(Y_e)_{ii'}\cr
&&-\beta^*(\psi_L^{e_i},\psi_R^{\nu_{i'}})(Y_\nu)_{ii'}+\alpha^*(\psi_L^{e_i},\psi_R^{e_{i'}})(Y_e)_{ii'}\cr
&&+\alpha(\psi_L^{u_i},\psi_R^{u_{i'}})(Y_u)_{ii'}+\beta(\psi_L^{u_i},\psi_R^{d_{i'}})(Y_d)_{ii'}\cr
&&-\beta^*(\psi_L^{d_i},\psi_R^{u_{i'}})(Y_u)_{ii'}+\alpha^*(\psi_L^{d_i},\psi_R^{d_{i'}})(Y_d)_{ii'}\big)+h.c.
\eea
where $q=\colvec{\alpha&\beta\cr -\beta^*&\alpha^*}$. We notice that the only form of $\eta_F$ which gives the correct kinetic term gives the correct sign for the Higgs term.

Finally, the neutrino mixing term is:
\bea
2S_f^{M}&=&( \psi,1\hat\otimes\sigma(z) \psi)_r\cr
&=&\sum_{p,p'}(\psi_R^p\otimes p\otimes R+ J_M\psi_R^p\otimes p\otimes \bar R,\cr
&&(\chi_M\otimes zM_0\otimes |\bar R\ket\bra R|-\chi_M\otimes z^*M_0^\dagger\otimes |R\ket\bra \bar R|)(\psi_R^{p'}\otimes p'\otimes R+ J_M\psi_R^{p'}\otimes p'\otimes \bar R))\cr
&=&\sum_{p,p'}(\psi_R^p\otimes p\otimes R,J_M\psi_R^{p'}\otimes z^*M_0^\dagger p'\otimes R)+\sum_{p,p'}(J_M\psi_R^p\otimes p\otimes \bar R,\psi_R^{p'}\otimes zM_0p'\otimes \bar R)\cr
&=&\sum_{p,p'}(\psi_R^p, J_M\psi_R^{p'})\bra p, z^*M_0^\dagger p'\ket+\sum_{p,p'}( J_M\psi_R^p,\psi_R^{p'})\bra p,zM_0p'\ket \cr
&=&\sum_{i,i'}(\psi_R^{\nu_i}, J_M\psi_R^{\nu_{i'}})\bra \nu_i, z^*m_0^\dagger \nu_i'\ket+\sum_{i,i'}( J_M\psi_R^{\nu_i},\psi_R^{{\nu_i}'})\bra \nu_i,zm_0\nu_i'\ket\cr
&=& {z}\sum_{i,i'}( J_M\psi_R^{\nu_i},\psi_R^{{\nu_i}'}) (m_0)_{ii'}+h.c.
\eea
To obtain the mass term, we expand around the minimum of the Higgses potential which is attained for $q=1$ and $z=1$ by construction. We thus see that $Y_\nu,Y_e,Y_u,Y_d$ are directly the physical  Dirac mass matrices, and $m_0$ is the Majorana mass matrix of the neutrinos. The singular values of these matrices are the  masses of the fermions.

\begin{rem}
The Connes-Lott and Spectral action here give different results.  With the latter, the Dirac operator entries $Y_x$ have to be redefined in terms of the physical mass matrices $m_x$, and must be supposed to be anti-hermitian (see \cite{VS} p 206). The interpretation is here more direct, but in return we do not get any relation between fermion masses as   in 12.1.3 in \cite{VS}.
\end{rem}

%


\section{Conclusion, Outlook}
In this paper, we have shown that a $U(1)$-extension of the SM, where the additional symmetry is broken by a new complex scalar field, comes out naturally of the algebraic background framework applied to the NCSM. It is a real strength of the NCG point of view that all the correct charges, signs and symmetries pop out by themselves by just ``turning a crank''. In particular, we have noted at several places that the single choice of $\eta_F$ makes plenty of independent signs right (fermionic kinetic terms and Yukawa couplings, symmetry of Majorana matrix $m_0$). But this choice is precisely the one which makes the finite background effectively Euclidean when it is combined with a quite complicated rule for graded tensor product of algebraic backgrounds,  modelled on Clifford algebras ! This adds to several other ``little miracles'' already well-known in the NCG approach such as the fact that the single KO-dimension in which fermion doubling can be solved is precisely the one in which the usual see-saw mechanism is possible. 
Even the Grassmann nature of the fermionic variables can be seen in a new light: at first we had left the two possibilities $s=\pm 1$ open, but eventually the choice $s=1$ led to an inconsistent sign for the kinetic term of the new Higgs.

However, the feeling that everything seems to fall in place, as intellectually satisfactory as it may be, is far from sufficient, and model has to be checked against experiment. As is apparent from \eqref{couplings1}, \eqref{couplings2}, \eqref{massbosons1} and \eqref{massbosons2},  the model presented here  makes some predictions at the unification scale. For instance, since $a$ is the sum of the squared Dirac masses of the fermions we obtain a relation between the mass of the $W$-bosons and the mass of the fermions:
\bea
M_W^2&=&{1\over 4}v^2g_w^2\cr
&=&{1\over 4}{1\over 32N}32 \tr(Y_e Y_e^\dagger+Y_\nu Y_\nu^\dagger+3M_u+3M_d)\cr
&=&{1\over 4N}\sum\mbox{ squared masses of fermions}
\eea
In particular for $N=3$, we obtain this bound for the mass of the top quark:
\be 
M_{\rm top}\le 2M_W
\ee
This prediction is different from the one obtained with the Spectral Action, which is $M_{\rm top}\le \sqrt{8/3}M_W$. The values obtained for the quartic couplings are quite striking: $\lambda_2$ is the square of the quotient of the standard deviation of the eigenvalues of $m_0m_0^\dagger$ by its mean, i.e., it is the relative standard deviation squared of the Majorana masses of the neutrinos ! The value for $\lambda_1$ is similar up to corrective terms coming from the angles $\theta_\ell$ and $\theta_q$. What remains to be done is to run down the coupling constants from the unification scale and obtain predictions for the masses of the Higsses and $Z'$ boson,  the value of the kinetic mixing. This will be the subject of a forthcoming paper.

\appendix
\section{Computation of $\hat J^1_{D_F}$}\label{appA}
A general element of $\hat \A_F$ can be written
\be
b= [\colvec{z_\nu &0\cr 0&z_e}\oplus \colvec{m_u&0\cr 0&m_d},\alpha\oplus\beta,\colvec{z_\nu & 0\cr 0& z_e^*}\oplus \colvec{n_u&0\cr 0&n_d},\gamma\oplus\delta]\otimes 1_N\label{elb}
\ee
where $\alpha,\gamma\in M_2(\CC)$ and $\beta,\delta\in M_2(\CC)\otimes M_3(\CC)$. We now want to calculate  the finite junk $2$-forms. For this we need to compute the elements of the form $b'[D_F,b]$. With obvious notations we have
\be
b'[D_F,b]=\colvec{0&b_R'[D_F,b]_{RL}&0&0\cr b_L'[D_F,b]_{LR}&0&0&0\cr 0&0&0&b_{\bar R}'[D_F,b]_{\bar R\bar L}\cr 0&0&b_{\bar L}'[D_F,b]_{\bar L\bar R}&0}\label{obvious1}
\ee
with 
\bea
b_R'[D_F,b]_{RL}&=&\colvec{z_\nu'&0\cr 0&z_e'}\left[\colvec{z_\nu&0\cr 0&z_e}Y_\ell^\dagger-Y_\ell^\dagger\alpha\right]\oplus\colvec{m_u'&0\cr 0&m_d'}\left[\colvec{m_u&0\cr 0&m_d}Y_q^\dagger-Y_q^\dagger\beta\right]\cr
b_L'[D_F,b]_{LR}&=&\alpha'\left[Y_\ell\colvec{z_\nu&0\cr 0&z_e}-\alpha Y_\ell\right]\oplus\beta'\left[Y_q\colvec{m_u&0\cr 0&m_d}-\beta Y_q\right]\cr
b_{\bar R}'[D_F,b]_{\bar R\bar L}&=&\colvec{z_\nu'&0\cr 0& (z_e')^*}\left[\colvec{z_\nu&0\cr 0& z_e^*}Y_\ell^T-Y_\ell^T\gamma\right]\oplus\colvec{n_u'&0\cr 0&n_d'}\left[\colvec{n_u&0\cr 0&n_d}Y_q^T-Y_q^T\delta\right]\cr
b_{\bar L}'[D_F,b]_{\bar L\bar R}&=&\gamma'\left[Y^*_\ell\colvec{z_\nu&0\cr 0&z_e^*}-\gamma Y^*_\ell\right]\oplus\delta'\left[Y^*_q\colvec{n_u&0\cr 0&n_d}-\delta Y^*_q\right]
\eea
where we have suppressed the $\otimes 1_N$. Now the $Y$ matrices are diagonal and act only on generations, hence they commute with the diagonal matrices acting trivially on generations. It follows that the leptonic and baryonic parts of $b_R'[D_F,b]_{RL}$ can be factorized on the left by $Y_\ell^\dagger$ and $Y_q^\dagger$ respectively, and similarly for the other matrix elements. We thus have
\be
\hat\Omega^1_{F}=\colvec{0&Y_\ell^\dagger M_2(\CC)\oplus Y_q^\dagger M_6(\CC)&0&0\cr
M_2(\CC)Y_\ell \oplus M_6(\CC) Y_q&0&0&0\cr
0&0&0&Y_\ell^T M_2(\CC)\oplus Y_q^T M_6(\CC)\cr
0&0&M_2(\CC) Y_\ell^*\oplus M_6(\CC) Y_q^*&0}\label{modformb}
\ee

Let $\Phi$ be a finite junk 2-form. It  is an element of the form $\Phi=\sum_i [D_F,b_i'][D_F,b_i]$ with $\sum_i b_i'[D_F,b_i]=0$. Let us write
\be
\Phi=\diag(\Phi_{RR}^\ell\oplus\Phi_{RR}^q,\Phi_{LL}^\ell\oplus\Phi_{LL}^q,\Phi_{\bar R\bar R}^\ell\oplus\Phi_{\bar R\bar R}^q,\Phi_{\bar L\bar L}^\ell\oplus\Phi_{\bar L\bar L}^q)\label{junkphi}
\ee
where
\bea
\Phi_{RR}^\ell&=&Y_\ell^\dagger\sum_i \left[\colvec{{z_\nu^i}'&0\cr 0&{z_e^i}'}- {\alpha^i}'\right]\left[\colvec{z_\nu^i&0\cr 0&z_e^i}-\alpha^i \right]Y_\ell,\cr
\Phi_{RR}^q&=&Y_q^\dagger\sum_i \left[\colvec{{m_u^i}'&0\cr 0&{m_d^i}'}-{\beta^i}'\right]\left[\colvec{m_u^i&0\cr 0&m_d^i}-\beta^i\right]Y_q
\eea
and so on, submitted to the conditions
\bea
Y_\ell^\dagger\sum_i\colvec{{z_\nu^i}'&0\cr 0&{z_e^i}'}\left[\colvec{z_\nu^i&0\cr 0&z_e^i}-\alpha^i\right]&=&0\cr
\sum_i{\alpha^i}'\left[\colvec{z^i_\nu&0\cr 0&z^i_e}-\alpha^i \right]Y_\ell&=&0\cr
\sum_i{\beta^i}'\left[\colvec{m_u^i&0\cr 0&m_d^i}-\beta^i \right]Y_q&=&0\cr
Y_q^\dagger\sum_i\colvec{{m_u^i}'&0\cr 0&{m_d^i}'}\left[\colvec{m_u^i&0\cr 0&m_d^i}-\beta^i\right]&=&0\cr
Y_\ell^T\sum_i\colvec{{z_\nu^i}'&0\cr 0&({ z_e^i}')^*}\left[\colvec{z_\nu^i&0\cr 0&(z_e^i)^*}-\gamma^i\right]&=&0\cr
\sum_i{\gamma^i}'\left[\colvec{z_\nu^i&0\cr 0& (z_e^i)^*}-\gamma^i \right]Y_\ell^*&=&0\cr
Y_q^T\sum_i\colvec{{n_u^i}'&0\cr 0&{n_d^i}'}\left[\colvec{n_u^i&0\cr 0&n_d^i}-\delta^i\right]&=&0\cr
\sum_i{\delta^i}'\left[\colvec{n_u^i&0\cr 0&n_d^i}-\delta^i \right]Y_q^*&=&0\label{junkconditions}
\eea
Thanks to the first four conditions, we easily obtain $\Phi_{RR}=0$, and we can prove similarly that $\Phi_{\bar R\bar R}=0$.
 
To deal with the $LL$-part, we first use the new variables $\tilde\alpha=\alpha-\colvec{z_\nu&0\cr 0&z_e}$. Thanks to the genericity hypothesis, the two first equations of \eqref{junkconditions} are equivalent to
\bea
\sum_i\colvec{{z_\nu^i}'&0\cr 0&{z_e^i}'} \tilde\alpha^i&=&0\cr
\sum_i{\tilde\alpha^i}{}'\tilde\alpha^i &=&0\label{eq59}
\eea
Rewriting $\Phi_{LL}^\ell$ with the new variables we obtain:
\bea
\Phi_{LL}^\ell&=&\sum_i \tilde{\alpha^i}{}' Y_\ell Y_\ell^\dagger\tilde\alpha^i\label{eq60}
\eea
Since we can always choose $z_\nu=z_e=0$ in the first condtion of \eqref{eq59}, the set of all elements in the form \eqref{eq60} submitted to the two conditions of \eqref{eq59} is the same as the set of the elements submitted only to the second condition. Hence, suppressing the tildes, we conclude that the $\Phi_{LL}^\ell$ part of the finite junk 2-forms is of the form
\be
\Phi_{LL}^\ell=\sum_i  {\alpha^i}{}' Y_\ell Y_\ell^\dagger \alpha^i,\mbox{ with }\sum_i{ \alpha^i}{}' \alpha^i=0\label{eq159}
\ee
In a completely similar manner, we obtain
\be
\Phi_{LL}^q=\sum_i  {\beta^i}{}' Y_q Y_q^\dagger \beta^i,\mbox{ with }\sum_i{ \beta^i}{}' \beta^i=0\label{eq160}
\ee
Now let us make the following general observation. Let $A$ be an $N$-dimensional unital $\RR$-algebra, let $r$ be an element of $A$ and define $m_r : A\otimes_\RR A\rightarrow A$ by $a\otimes b\mapsto arb$. What we are looking for in the case of $\Phi_{LL}^\ell$ is $m_r(\ker (m_1))$, with $A=M_2(\CC)$ and $r=Y_\ell Y_\ell^\dagger$. First we know that $\ker(m_1)$ is generated as a vector space over $\RR$ by the elements of the form $x\otimes y-xy\otimes 1$ (Indeed, $a=\sum x_i\otimes y_i$ with $\sum x_iy_i=0$ can be rewritten $a=\sum (x_i\otimes y_i-x_iy_i\otimes 1)$). Thus, if $(x_i)_{1\le i\le N}$ is a $\RR$-basis of $A$, then the set $x_i\otimes x_j-x_ix_j\otimes 1$ is  generating for $\ker(m_1)$. Hence, if we suppose that $x_N=1$, we conclude that $(x_i[r,x_j])_{1\le i\le N \atop 1\le j< N}$ is a generating set for $m_r(\ker (m_1))$. (Let us remark that this set has (at most) $N(N-1)$ elements, and this is precisely the dimension of $\ker(m_1)$ since $m_1$ is surjective. Thus $(x_i\otimes x_j-x_ix_j\otimes 1)_{1\le i\le N \atop 1\le j< N}$ is a basis of $\ker(m_1)$.)

Applying the above observation, we obtain that $\Phi_{LL}^q$ is a general linear combination of $x_i[ Y_\ell Y_\ell^\dagger,x_j]$ where $x_i$ and $x_j$ run through a $\RR$-basis of $M_2(\CC)$. Considering that basis $(E_{ij},iE_{ij})$, where $E_{ij}$ the elementary matrix with $1$ in position $(i,j)$, we easily obtain that
\be
\Phi_{LL}^\ell=\alpha'\otimes(Y_\nu Y_\nu^\dagger-Y_e Y_e^\dagger)
\ee
where $\alpha'$ is any element of $M_2(\CC)$ and $Y_\nu Y_\nu^\dagger=Y_\nu Y_\nu^\dagger$, $Y_e Y_e^\dagger=Y_e Y_e^\dagger$. Similarly we have
\bea
\Phi_{LL}^q&=&\beta'\otimes (Y_u Y_u^\dagger-Y_d Y_d^\dagger),\ \beta'\in M_2(\CC)\otimes M_3(\CC)\cr
\Phi_{\bar L\bar L}^\ell&=&\gamma'\otimes( Y_\nu^* Y_\nu^T-Y_e^* Y_e^T),\ \gamma'\in M_2(\CC)\cr
\Phi_{\bar L\bar L}^q&=&\delta'\otimes ( Y_u^* Y_u^T- Y_d^* Y_d^T),\ \delta'\in M_2(\CC)\otimes M_3(\CC)
\eea
An element of the total junk $\hat\J_1$ is a function with values in $\hat \A_F+\hat \J_{D_F}^1$, that is, with values of the form
\bea
[\colvec{z_\nu &0\cr 0&z_e}\otimes 1_N\oplus \colvec{m_u&0\cr 0&m_d}\otimes 1_N,(\alpha\otimes 1_N+\alpha'\otimes(Y_\nu Y_\nu^\dagger-Y_e Y_e^\dagger))\oplus(\beta\otimes 1_N+\beta'\otimes (Y_u Y_u^\dagger-Y_d Y_d^\dagger)),\cr
\colvec{z_\nu & 0\cr 0& z_e^*}\otimes 1_N\oplus \colvec{n_u&0\cr 0&n_d}\otimes 1_N,(\gamma\otimes 1_N+\gamma'\otimes(Y_\nu ^*Y_\nu^T- Y_e^* Y_e^T))\oplus(\delta\otimes 1_N+\delta'\otimes ( Y_u^* Y_u^T- Y_d^* Y_d^T)]
\eea
We now look for an orthonormal basis of $\hat \A_F+\hat \J_{D_F}^1$. For this, let us consider orthonormal $\RR$-bases $(\alpha_i)_{1\le i\le 8}$ of $M_2(\CC)$ and $(\beta_i)_{1\le i\le 9}$ of $M_3(\RR)$. We also consider the family $\lambda_i\oplus\lambda_i'\in M_2(\CC)\oplus M_2(\CC)$ defined by
\be
\lambda_{1/2/3/4}=\colvec{1&0\cr 0&0}, \colvec{i&0\cr 0&0},\colvec{0&0\cr 0&1},\colvec{0&0\cr 0&i},\mbox{ respectively,}
\ee
and 
\be
\lambda_{1/2/3/4}'=\colvec{1&0\cr 0&0}, \colvec{i&0\cr 0&0},\colvec{0&0\cr 0&1},\colvec{0&0\cr 0&-i},\mbox{ respectively.}
\ee
\begin{lemma} The family consisting of
\begin{enumerate}
\item $Z_i={1\over\sqrt{2N}}\diag(\lambda_i\oplus 0,0,\lambda_i'\oplus 0,0)\otimes 1_N$,
\item $M_{ij}={1\over \sqrt{N}}\diag(0\oplus \lambda_i\otimes \beta_j\otimes 1_N,0,0,0)$,
\item $A_i={1\over \sqrt{N}}\diag(0,\alpha_i\otimes 1_N\oplus 0,0,0)$
\item $B_{ij}={1\over \sqrt{N}}\diag(0,0\oplus \alpha_i\otimes \beta_j\otimes 1_N,0,0)$,
\item $A_i'={1\over k_\ell}\diag(0,\alpha_i\otimes T_\ell\oplus 0,0,0)$,
\item $B_{ij}'={1\over k_q}\diag(0,0\oplus \alpha_i\otimes \beta_j\otimes T_q,0,0)$,
\end{enumerate}
where $T_\ell={\widetilde{Y_\nu Y_\nu^\dagger}-\widetilde{Y_e Y_e^\dagger}}$, $T_q={\widetilde{Y_\nu Y_\nu^\dagger}-\widetilde{Y_e Y_e^\dagger}}$, $k_\ell=\|\widetilde{Y_\nu Y_\nu^\dagger}-\widetilde{Y_e Y_e^\dagger}\|_{KS}=(\tr T_\ell^2)^{1/2}$, and $k_q=\|\widetilde{Y_u Y_u^\dagger}-\widetilde{Y_d Y_d^\dagger}\|_{KS}=(\tr T_q^2)^{1/2}$, together with $M_{ij}^o, A_i^o$, etc., is an orthonormal basis of $\hat \A_F+\hat \J_{D_F}^1$ for the Krein-Schmidt product.
\end{lemma}

\section{Computation of $(\hat\J^1_F)^\ext$}\label{appB}
A general element of $\hat \A_F^\ext$ is of the form
\be
b=[\colvec{z_\nu &0\cr 0&z_e}\oplus \colvec{m_u&0\cr 0&m_d},\alpha\oplus\beta,\colvec{z_{\bar \nu} & 0\cr 0&  z_{\bar e}}\oplus \colvec{m_{\bar u}&0\cr 0&m_{\bar d}},\gamma\oplus\delta]\otimes 1_N\label{bext}
\ee
where now $z_\nu,z_e,z_{\bar \nu}, z_{\bar e}$ are four independent complex numbers. Hence  $\hat \A_F^\ext=\hat \A_F\oplus \CC p_{\nu_{\bar R}}\oplus \CC p_{ e_{\bar R}}$, though the decomposition  $\hat \A_F^\ext=\hat \A_F\oplus \CC p_{\bar \nu}\oplus \CC p_{\bar e}$, with $p_{\bar \nu}=[0,0,\colvec{1&0\cr 0&0}\oplus 0,\colvec{1&0\cr 0&0}\oplus 0]\otimes 1_N$ and  $p_{\bar e}=[0,0,\colvec{0&0\cr 0&1}\oplus 0,\colvec{0&0\cr 0&1}\oplus 0]\otimes 1_N$ will be more useful in the sequel.

Repeating the calculation \eqref{obvious1} with now $b,b'\in  \hat \A_F^\ext$, we find
\be
b'[D_F,b]=\colvec{0&b_R'[D_F,b]_{RL}&b_R'[D_F,b]_{R\bar R}&0\cr b_L'[D_F,b]_{LR}&0&0&0\cr b_{\bar R}'[D_F,b]_{\bar RR}&0&0&b_{\bar R}'[D_F,b]_{\bar R\bar L}\cr 0&0&b_{\bar L}'[D_F,b]_{\bar L\bar R}&0}\label{obvious2}
\ee
with 
\bea
b_R'[D_F,b]_{RL}&=&Y_\ell^\dagger\colvec{z_\nu'&0\cr 0&z_e'}\left[\colvec{z_\nu&0\cr 0&z_e}-\alpha\right]\oplus Y_q^\dagger\colvec{m_u'&0\cr 0&m_d'}\left[\colvec{m_u&0\cr 0&m_d}- \beta\right]\cr
b_L'[D_F,b]_{LR}&=&\alpha'\left[\colvec{z_\nu&0\cr 0&z_e}-\alpha \right]Y_\ell\oplus\beta'\left[\colvec{m_u&0\cr 0&m_d}-\beta \right]Y_q\cr
b_{\bar R}'[D_F,b]_{\bar R\bar L}&=&Y_\ell^T\colvec{z_{\bar \nu}'&0\cr 0& z_{\bar e}'}\left[\colvec{z_{\bar \nu}&0\cr 0& z_{\bar e}}-\gamma\right]\oplus Y_q^T\colvec{m_{\bar u}'&0\cr 0&m_{\bar d}'}\left[\colvec{m_{\bar u}&0\cr 0&m_{\bar d}}-\delta\right]\cr
b_{\bar L}'[D_F,b]_{\bar L\bar R}&=&\gamma'\left[\colvec{z_{\bar \nu}&0\cr 0& z_{\bar e}}-\gamma \right]Y^*_\ell\oplus\delta'\left[\colvec{m_{\bar u}&0\cr 0&m_{\bar d}}-\delta \right]Y^*_q\cr
b_R'[D_F,b]_{R\bar R}&=&sz_\nu'(z_{\bar \nu}-z_\nu)M_0^\dagger\oplus 0\cr
b_{\bar R}'[D_F,b]_{\bar R R}&=&z_{\bar \nu}'(z_\nu-z_{\bar \nu})M_0\oplus 0
\eea
Hence we see that
\be
(\hat\Omega^1_{F})^\ext=\colvec{0&Y_\ell^\dagger M_2(\CC)\oplus Y_q^\dagger M_6(\CC)&\CC M_0^\dagger&0\cr
M_2(\CC)Y_\ell \oplus M_6(\CC) Y_q&0&0&0\cr
\CC M_0&0&0&Y_\ell^T M_2(\CC)\oplus Y_q^T M_6(\CC)\cr
0&0&M_2(\CC) Y_\ell^*\oplus M_6(\CC) Y_q^*&0}\label{modformbext}
\ee
%
We notice that $[D_F,p_{\bar e}]=0$ and $[D_F,p_{\bar \nu}]=\colvec{0&0&M^\dagger&0\cr 0&0&0&0\cr -M&0&0&0\cr 0&0&0&0}$, and from \eqref{modformb} and \eqref{bext}   we get $\hat\A_F^\ext\hat\Omega^1_{F}=\hat\Omega^1_{F}$. Let $a',b'\in \hat \A_F^\ext$. Using $\hat \A_F^\ext=\hat \A_F\oplus \CC p_{\bar \nu}\oplus \CC p_{\bar e}$, we write $b'=b+zp_{\bar \nu}+wp_{\bar e}$, so that
\bea
a'[D_F,b']&=&a'[D_F,b+zp_{\bar \nu}+wp_{\bar e}]\cr
&=&a'[D_F,b]+a'[D_F,zp_{\bar \nu}]\nonumber
\eea
and we thus obtain $(\hat\Omega^1_{F})^\ext=\hat\Omega^1_{F}\oplus\Omega^1_\sigma$.

Let us compute the finite junk 2-forms. With obvious notations let us consider a vanishing sum
\bea
0=\sum_i a_i'[D_F,b_i+z_ip_{\bar \nu}+w_ip_{\bar e}]&=&\sum_i a_i'[D_F,b_i]+a_i'[D_F,z_ip_{\bar\nu}]\cr
&\in&\hat\Omega^1_{F}\oplus\Omega^1_\sigma\nonumber
\eea
with $a_i'\in \hat \A_F^\ext$ and $b_i\in \hat \A_F$. Since the two terms must vanish,  the bimodule of junk 2-forms is the direct sum of two types of elements:
\begin{enumerate}
\item $\rho_1=\sum[D_F,a_i'][D_F,b_i]$ with $\sum a_i'[D_F,b_i]=0$,
\item $\rho_2=\sum[D_F,a_i'][D_F,z_ip_{\bar\nu}]$, with $\sum a_i'[D_F,z_i p_{\bar \nu}]=0$,
\end{enumerate}
An element of the first type is of the form $\Phi+\Phi'$, where $\Phi$ is
\be
\Phi=\diag(\Phi_{RR}^\ell\oplus\Phi_{RR}^q,\Phi_{LL}^\ell\oplus\Phi_{LL}^q,\Phi_{\bar R\bar R}^\ell\oplus\Phi_{\bar R\bar R}^q,\Phi_{\bar L\bar L}^\ell\oplus\Phi_{\bar L\bar L}^q)
\ee
as in \eqref{junkphi}, except that the complex numbers on the particle and anti-particle blocks are now independent, and
\be
\Phi'=\sum_i\colvec{0&0&0&[D,a_i']_{R\bar R}[D,b_i]_{\bar R\bar L}\cr 0&0&0&0\cr 0&[D,a_i']_{\bar R R}[D,b_i]_{RL}&0&0\cr 0&0&0&0}
\ee
The analysis of the $\Phi$ part stays the same as before: some complex numbers are now independent but since they were never used together, nothing changes. The components of the $\Phi'$ part are
\bea
[D_F,a_i']_{R\bar R}[D_F,b_i]_{\bar R\bar L}&=&sM_0^\dagger Y_\ell^T\sum_i\colvec{(z_{\bar \nu}^i)'-(z_\nu^i)'&0\cr 0&(z_{\bar e}^i)'-(z_e^i)'} \left[\colvec{z_{ \nu}^i&0\cr 0& (z_{e}^i)^*} -\gamma_i\right]\cr
[D,a_i']_{\bar R R}[D,b_i]_{RL}&=&-M_0 Y_\ell^\dagger\sum_i\colvec{(z_{\bar \nu}^i)'-(z_\nu^i)'&0\cr 0&(z_{\bar e}^i)'-(z_e^i)'} \left[\colvec{z_\nu^i&0\cr 0&z_e^i}-\alpha_i\right]\nonumber
\eea
submitted to the  conditions
\bea
\sum_i\colvec{{z_{\bar \nu}^i}'&0\cr 0&{z_{\bar e}^i}'}\left[\colvec{z_{ \nu}^i&0\cr 0&(z_{ e}^i)^*}-\gamma^i\right]&=&0\cr
\sum_i{\gamma^i}'\left[\colvec{z_\nu^i&0\cr 0& (z_e^i)^*}-\gamma^i \right]&=&0
\eea
We finally find that $\Phi'$ has the form
\be
\Phi'= \colvec{0&0&0&M_0^\dagger Y_\ell^T\gamma_1\cr 0&0&0&0\cr 0& M_0Y_\ell^\dagger \gamma_2&0&0\cr 0&0&0&0}
\ee
where $\gamma_1,\gamma_2\in M_2(\CC)$ are arbitrary (this can be seen for instance by taking special elements of the form $[D_F,a'][D_F,b]$ with $a'=[\colvec{z_\nu'&0\cr 0&z_e'}\oplus 0,0,0,0]$ and $b=[0,0,0,\gamma\oplus 0]$). Let us now look at the elements of the second type. Writing the $\CC^4$ factor of $\hat\A_F^\ext$ in the form $\CC_\nu\oplus\CC_e\oplus \CC_{\bar \nu}\oplus\CC_{\bar e}$, let us write $(z_\nu^i)'$ and $(z_{\bar \nu}^i)'$ the coordinates of $a_i'$ in the $\CC_\nu$ and $\CC_{\bar \nu}$ factors, respectively. The condition $\sum a_i'[D_F,z_i p_{\bar \nu}]=0$ is the equivalent to $\sum_i (z_\nu^i)'z_i=\sum_i (z_{\bar \nu}^i)'z_i=0$. The corresponding element $\rho_2$ is
\be
\rho_2=\sum\colvec{-[D_F,a_i']_{R\bar R}z_iM_0& 0&0&0\cr 0&0&[D_F,a_i']_{LR}z_iM_0^\dagger&0\cr 0&0&[D_F,a_i']_{\bar RR}z_iM_0^\dagger&0\cr -[D_F,a_i']_{\bar L\bar R}z_iM_0&0&0&0} 
\ee
with
\bea
-[D_F,a_i']_{R\bar R}z_iM_0&=&-\sum ((z_{\bar \nu}^i)'-(z_\nu^i)')z_iM_0^\dagger M_0=0\cr
[D_F,a_i']_{\bar RR}z_iM_0^\dagger&=&\sum ((z_\nu^i)'-(z_{\bar \nu}^i)')z_iM_0M_0^\dagger=0\cr
[D_F,a_i']_{LR}z_iM_0^\dagger&=&\sum\left[Y_\ell\colvec{(z_\nu^i)'&0\cr 0&(z_e^i)'}-\alpha_i' Y_\ell\right]z_iM_0^\dagger=-\sum\alpha_i'z_i Y_\ell M_0^\dagger\cr
-[D_F,a_i']_{\bar L\bar R}z_iM_0&=&-\sum\left[Y^*_\ell\colvec{(z_{\bar \nu}^i)'&0\cr 0& (z_{\bar e}^i)'}-\gamma_i'Y^*_\ell\right]z_iM_0=-\sum\gamma_i'z_i Y^*_\ell M_0\cr
\eea
Thus, $\rho_2$ has the form
\be
\rho_2=\colvec{0& 0&0&0\cr 0&0& \alpha_1Y_\ell M_0^\dagger&0\cr 0&0&0&0\cr \alpha_2 Y^*_\ell M_0&0&0&0} 
\ee
where $\alpha_1,\alpha_2$ are any $2\times 2$ matrices (whose first columns only count). Hence the junk 2-forms of $\hat\A_F^\ext$ are the junk 2-forms of $\hat\A_F$ plus an extra antidiagonal part of the form:
\be
\Phi'+\rho_2= \colvec{0&0&0&M_0^\dagger Y_\ell^T\gamma_1\cr 0&0& \alpha_1Y_\ell M_0^\dagger&0\cr 0& M_0Y_\ell^\dagger \gamma_2&0&0\cr \alpha_2 Y^*_\ell M_0&0&0&0}\label{antidiagjunk}
\ee
We call ${\cal M}$ the module (isomorphic to $M_2(\CC)^4$) of these antidiagonal junk forms. With this notation, the total junk 2-forms of $\hat\B^\ext$ are functions with values in $\hat \A_F^\ext+\hat\J^1_F+{\cal M}$, that is, with values of the form
\bea
[\colvec{z_\nu &0\cr 0&z_e}\otimes 1_N\oplus \colvec{m_u&0\cr 0&m_d}\otimes 1_N,(\alpha\otimes 1_N+\alpha'\otimes(Y_\nu Y_\nu^\dagger-Y_e Y_e^\dagger))\oplus(\beta\otimes 1_N+\beta'\otimes (Y_u Y_u^\dagger-Y_d Y_d^\dagger)),\cr
\colvec{z_{\bar\nu} & 0\cr 0&z_{\bar e}}\otimes 1_N\oplus \colvec{m_{\bar u}&0\cr 0&m_{\bar d}}\otimes 1_N,(\gamma\otimes 1_N+\gamma'\otimes(Y_\nu^* Y_\nu^T-Y_e^* Y_e^T))\oplus(\delta\otimes 1_N+\delta'\otimes ( Y_u^* Y_u^T- Y_d^* Y_d^T)]
\eea
plus an additional antidiagonal part.  The basis is changed accordingly, and we obtain
\begin{lemma}\label{basejunk2} The family consisting of
\begin{enumerate}
\item $Z_i={1\over\sqrt{N}}\diag(\lambda_i\oplus 0,0,0,0)\otimes 1_N$,
\item $M_{ij}={1\over \sqrt{N}}\diag(0\oplus \lambda_i\otimes \beta_j\otimes 1_N,0,0,0)$,
\item $A_i={1\over \sqrt{N}}\diag(0,\alpha_i\otimes 1_N\oplus 0,0,0)$
\item $B_{ij}={1\over \sqrt{N}}\diag(0,0\oplus \alpha_i\otimes \beta_j\otimes 1_N,0,0)$,
\item $A_i'={1\over k_\ell}\diag(0,\alpha_i\otimes T_\ell\oplus 0,0,0)$,
\item $B_{ij}'={1\over k_q}\diag(0,0\oplus \alpha_i\otimes \beta_j\otimes T_q,0,0)$,
\end{enumerate}
 together with $Z_i^o,M_{ij}^o, A_i^o$, etc., is an orthonormal basis of $\hat\A_F^\ext+\hat\J^1_F$ for the Krein-Schmidt product. 
\end{lemma}
Note that the antidiagonal part is orthogonal to  $\hat\A_F^\ext+\hat\J^1_F$ and its basis will not matter.

\begin{rem} If $s=-1$ the Krein-Schmidt product is positive definite on matrices which have a diagonal $+$ antidiagonal form, hence on $\hat\A_F^\ext+\hat{\J^1_F}^{\rm ext}$. If $s=1$ it has neutral signature on these matrices. In both cases the Krein-Schmidt is non-degenerate on $\hat\A_F^\ext+\hat{\J^1_F}^{\rm ext}$ as required.
\end{rem}
\section{Projection of the Higgs and $\sigma$-curvature}\label{projectionHiggs}
We first compute the finite curvature of $\Phi(q')$. For this, we remark that we can write
\be 
\Phi(q')=\pi_F^\ext(1,0,0,0)[D_F,\pi_F(0,(q')^\dagger,0,0)]+\pi_F^\ext(0,1,0,0)[D_F,\pi_F^\ext(0,-q',0,0)]
\ee
We infer from this the finite differential:
\bea
d_{D_F}\Phi(q')&=&[D_F,\pi_F^\ext(1,0,0,0)][D_F,\pi_F^\ext(0,(q')^\dagger,0,0)]+[D_F,\pi_F^\ext(0,1,0,0)][D_F,\pi_F^\ext(0,-q',0,0)]\cr
&=&\colvec{
-Y_0^\dagger (\tilde q'+\tilde{q'}^\dagger)Y_0 &0&0&0\cr
0&-Y_0Y_0^\dagger( \tilde{q'}+\tilde{q'}^\dagger)&0&0\cr
0&-M_0Y_\ell^\dagger \tilde{q'}^\dagger &0&0\cr 
0&0&0&0
}\cr
&=&-2\reel(q')[Y_0^\dagger Y_0,Y_0Y_0^\dagger,0,0]+{\rm junk}
\eea
Now we also have
\bea
\Phi(q')^2&=&-[Y_0^\dagger \tilde{q'}^\dagger \tilde{q'}Y_0,\tilde q' Y_0Y_0^\dagger\tilde{q'}^\dagger,0,0]\cr
&=&-|q'|^2[Y_0^\dagger Y_0,Y_0Y_0^\dagger,0,0]+{\rm junk}
\eea
where in the last line we have used the following trick: $q'{q'}^\dagger-|q'|^21_2=0\Rightarrow [0,\tilde q' Y_0Y_0^\dagger\tilde{q'}^\dagger- Y_0Y_0^\dagger\tilde q' \tilde{q'}^\dagger,0,0]\in (\hat\J^1_{D_F})^\ext$, from \eqref{eq159} and \eqref{eq160}. We thus obtain
\be 
\rho_{\rm Higgs}=-(|q|^2-1)^2[Y_0^\dagger Y_0,Y_0Y_0^\dagger,0,0]:=-(|q|^2-1)^2\phi
\ee
We compute $P(\phi)$ by the formula $P(\phi)=\phi-\sum_i (e_i,\phi)_\RR e_i$, where $e_i$ runs through the basis of lemma \ref{basejunk2}, and $(.,.)_\RR$ is the real Krein-Schmidt product which is a scalar product in restriction to block-diagonal matrices. We find:
\bea
P(\phi)&=&\phi-\big[{1\over\sqrt{N}}(\tr Y_\nu Y_\nu^\dagger)Z_1+{1\over\sqrt{N}}(\tr Y_e Y_e^\dagger)Z_3+\sqrt{3\over N}\tr(Y_u Y_u^\dagger)M_{11}\cr
&&+\sqrt{3\over N}\tr(Y_d Y_d^\dagger)M_{31}+{1\over\sqrt{2N}}\tr(Y_\nu Y_\nu^\dagger+Y_e Y_e^\dagger)A_1+{1\over \sqrt{2N}}\tr(Y_\nu Y_\nu^\dagger-Y_e Y_e^\dagger)A_2\cr
&&+\sqrt{3\over 2N}\tr(Y_u Y_u^\dagger+Y_d Y_d^\dagger)B_{11}+\sqrt{3\over 2N}\tr(Y_u Y_u^\dagger-Y_d Y_d^\dagger)B_{21}\cr
&&+{1\over k_\ell\sqrt{2}}(Y_\nu Y_\nu^\dagger+Y_e Y_e^\dagger,T_\ell)_\RR A_1'+{1\over k_\ell\sqrt{2}}(Y_\nu Y_\nu^\dagger-Y_e Y_e^\dagger,T_\ell)_\RR A_2'\cr
&&+{1\over k_q}\sqrt{3\over 2N}(Y_u Y_u^\dagger+Y_d Y_d^\dagger,T_q)_\RR B_{11}'+{1\over k_q}\sqrt{3\over 2N}(Y_u Y_u^\dagger-Y_d Y_d^\dagger,T_q)_\RR B_{21}'\big]\nonumber
\eea
Replacing the basis element with their expression, we find that
\be 
P(\rho_{\rm Higgs})=-(|q|^2-1)^2[C_1,C_2,0,0] 
\ee
where
\bea
C_1&=&\colvec{\widetilde{Y_\nu^\dagger Y_\nu}&0\cr 0&\widetilde{Y_e^\dagger Y_e}}\oplus 1_3\otimes\colvec{ \widetilde{Y_u^\dagger Y_u}&0\cr 0&\widetilde{Y_d^\dagger Y_d}}\cr
C_2&=&\colvec{\widetilde{Y_\nu Y_\nu^\dagger}-{1\over k_\ell^2}(T_\ell, Y_\nu Y_\nu^\dagger)_\RR T_\ell&0\cr 0&\widetilde{Y_e Y_e^\dagger}-{1\over k_\ell^2}(T_\ell, Y_e Y_e^\dagger)_\RR T_\ell}\cr
&&\oplus 1_3\otimes\colvec{\widetilde{Y_u Y_u^\dagger}-{1\over k_q^2}(T_q, Y_u Y_u^\dagger)_\RR T_q&0\cr 0&\widetilde{Y_d Y_d^\dagger}-{1\over k_q^2}(T_q, Y_d Y_d^\dagger)_\RR T_q}
\eea
Let us now calculate the projection of $\rho_\sigma$. To compute the finite differential of $\sigma(z)$, we use the decomposition\footnote{Note that this decomposition is meaningful in $\hat \Omega^1_\ext$ only.} (see \eqref{finitediffsigma}):
\bea
\sigma(x+iy)&=&{1\over 2}(xt_{B-L}[D_F,t_{B-L}]-y[D_F,t_{B-L}])
\eea
Thus
\bea
\rho_\sigma&=&{1\over 2}x  [D_F,t_{B-L}]^2+\sigma(z')^2\cr
&=&s(|z|^2-1)\left[M_0^\dagger M_0\oplus 0,0,M_0M_0^\dagger\oplus 0,0\right]\cr
&=&s(|z|^2-1)\left[\colvec{m_0^\dagger m_0&0\cr 0&0}\oplus 0,0,\colvec{m_0m_0^\dagger &0\cr 0&0}\oplus 0,0\right]
\eea
which is remarkably similar to the Higgs curvature. The projection is easy to calculate, as this matrix is orthogonal to every basis element except $Z_1$ and $Z_1^o$. It just remove the trace of $M_0^\dagger M_0$ and $M_0M_0^\dagger$. We thus have:
\bea
P(\rho_\sigma)&=&s(|z|^2-1)\left[\colvec{\widetilde{m_0^\dagger m_0}&0\cr 0&0}\oplus 0,0,\colvec{\widetilde{m_0m_0^\dagger} &0\cr 0&0}\oplus 0,0\right]\cr
&:=&s(|z|^2-1)[D_1,0,D_1^*,0]
\eea
where we have used $(m_0^\dagger m_0)^*=m_0^Tm_0^*=sm_0m_0^*=m_0m_0^\dagger$. The last computation we need is the Krein-Schmidt norm of the projection of the curvature. We have:
\bea
\reel \tr(P(\rho_{\rm Higgs}+\rho_{\rm Higgs}^o+\rho_\sigma)^2)&=&\reel \tr (-(|q|^2-1)[C_1,C_2,C_1^*,C_2^*]\cr
&&+s(|z|^2-1)[D_1,0,D_1,0])^2)\cr
&=&(|q|^2-1)^2\reel \tr(C_1^2+C_2^2+(C_1^*)^2+(C_2^*)^2)\cr
&&+(|z|^2-1)^2\reel\tr(D_1^2+(D_1^*)^2)\cr
&&-2s(|z|^2-1)(|q|^2-1)\reel\tr(C_1D_1+C_1^*D_1^*)\cr
&=&2(|q|^2-1)^2 \tr(C_1^2+C_2^2)+2(|z|^2-1)^2\tr(D_1^2)\cr
&&-4s(|z|^2-1)(|q|^2-1)\reel\tr(C_1D_1)\cr
&:=&2V_0(|q|^2-1)^2+2W_0(|z|^2-1)^2-4sK(|z|^2-1)(|q|^2-1)\cr
\label{normesq}
\eea
One then obtains
\bea
V_0&=&\tr(C_1^2+ C_2^2)\cr
&=&\|\widetilde{Y_\nu Y_\nu^\dagger}\|^2+\|\widetilde{Y_e Y_e^\dagger}\|^2+3\|\widetilde{Y_u Y_u^\dagger}\|^2+3\|\widetilde {Y_d Y_d^\dagger}\|^2\cr
&&+2{\|\widetilde{Y_\nu Y_\nu^\dagger}\|^2\|\widetilde{Y_e Y_e^\dagger}\|^2\over \|\widetilde{Y_\nu Y_\nu^\dagger}-\widetilde {Y_e Y_e^\dagger}\|^2}\sin^2(\theta_\ell)+6{\|\widetilde {Y_u Y_u^\dagger}\|^2\|\widetilde{Y_d Y_d^\dagger}\|^2\over \|\widetilde{Y_u Y_u^\dagger}-\widetilde{Y_d Y_d^\dagger}\|^2}\sin^2(\theta_q)\cr
W_0&=&\tr(D_1^2)=\|\widetilde {m_0m_0^\dagger}\|^2\cr
K&=&\reel \tr(C_1D_1)=\reel\tr(\widetilde{Y_\nu^\dagger Y_\nu}\widetilde{m_0^\dagger m_0})
\eea
where the angles $\theta_\ell$ and $\theta_q$ are defined up to sign by
\bea
(\widetilde{Y_\nu Y_\nu^\dagger},\widetilde{Y_e Y_e^\dagger})_\RR=\|\widetilde{Y_\nu Y_\nu^\dagger}\|\|\widetilde{Y_e Y_e^\dagger}\|\cos(\theta_\ell)\cr
(\widetilde{Y_u Y_u^\dagger},\widetilde{Y_d Y_d^\dagger})_\RR=\|\widetilde{Y_u Y_u^\dagger}\|\|\widetilde{Y_d Y_d^\dagger}\|\cos(\theta_q).
\eea
\bibliographystyle{unsrt}
\bibliography{../generalbib/SSTbiblio}
\end{document}